\documentclass[twocolumn]{autart}
\usepackage{graphicx}
\usepackage{amsmath,amssymb,amscd}
\usepackage{bm}
\usepackage{float}
\usepackage{boondox-cal}
\usepackage{threeparttable}
\usepackage{color}
\usepackage{hyperref}
\usepackage{stfloats}
\usepackage{subfigure}
\newtheorem{theorem}{Theorem}

\newtheorem{lemma}{Lemma}

\newtheorem{assumption}{Assumption}
\newtheorem{problem}{\indent Problem}
\begin{document}

\begin{frontmatter}

\title{  On the regularization and optimization in quantum detector tomography \thanksref{footnoteinfo}} 

\thanks[footnoteinfo]{ This research was supported by the National Natural Science Foundation
	of China (62173229, 12288201), the Australian Research Council's Future Fellowship funding scheme under
	Project FT220100656, the Centres of Excellence
	under Grant CE170100012, and U.S. Office of Naval Research Global under Grant N62909-19-1-2129. }

\author[1,2]{Shuixin Xiao}\ead{xiaoshuixin@sjtu.edu.cn},    
\author[3,4]{Yuanlong Wang}\ead{wangyuanlong@amss.ac.cn}, 
\author[1,a]{Jun Zhang}\ead{zhangjun12@sjtu.edu.cn},
\author[2,a]{Daoyi Dong}\ead{daoyidong@gmail.com},             
\author[2,5]{Shota Yokoyama}\ead{s.yokoyama@adfa.edu.au},  
\author[6]{Ian R. Petersen}\ead{i.r.petersen@gmail.com},  
\author[2,5]{Hidehiro Yonezawa}\ead{h.yonezawa@adfa.edu.au}  
\thanks[a]{Corresponding author.}

\address[1]{University of Michigan – Shanghai Jiao Tong University Joint Institute, Shanghai Jiao Tong
	University, Shanghai 200240, China}  
\address[2]{School of Engineering and Information Technology, University of New South Wales, Canberra ACT 2600, Australia} 
\address[3]{Key Laboratory of Systems and Control, Academy of Mathematics and Systems Science, Chinese Academy of Sciences, Beijing 100190, China} 
\address[4]{Centre for Quantum Computation and Communication Technology (Australian Research Council), Centre for Quantum Dynamics, Griffith University, Brisbane, Queensland 4111, Australia}
  \address[5]{Centre for Quantum Computation and Communication Technology, Australian Research Council, Canberra, ACT 2600, Australia}                                          
\address[6]{School of
	Engineering, Australian
	National University, Canberra, ACT 2601, Australia}


\begin{abstract}  
Quantum detector tomography (QDT) is a fundamental technique for calibrating quantum devices and  performing quantum engineering tasks. In this paper, we utilize regularization to improve the QDT accuracy whenever  the probe states are informationally complete or informationally incomplete.  In the informationally complete scenario, without regularization, 
we optimize the resource (probe state) distribution by converting it to a semidefinite programming problem.  Then in both the informationally complete and informationally incomplete scenarios,   we discuss different regularization forms and  prove the mean squared error scales as $ O(1/{N}) $ or tends to a constant with $ N $ state copies under the static assumption. We also characterize the ideal best regularization for the identifiable parameters, accounting for both the informationally complete and informationally incomplete scenarios. Numerical examples demonstrate the effectiveness of different regularization forms and a quantum optical experiment test shows that a suitable regularization form can reach a reduced mean squared error.
\end{abstract}

\end{frontmatter}

\endNoHyper

\section{Introduction}
In the past decades, significant progress has been achieved in a variety of fields of quantum science and technology, including quantum computation \cite{DiVincenzo255}, quantum communication \cite{qci} and quantum sensing \cite{RevModPhys.89.035002}. In these applications, it is often necessary to develop efficient estimation methods to acquire information about quantum systems and quantum system identification has attracted wide attention \cite{D1,NURDIN2022,7130587}. In quantum estimation and quantum system identification, a common and essential step is to perform measurement on the quantum system of interest. Quantum detector tomography (QDT), as the standard technique to characterize an unknown measurement process, is fundamental for device benchmarking and 
subsequent tasks such as quantum state tomography (QST) \cite{Qi2013,Hou2016,MU2020108837}, quantum Hamiltonian identification \cite{8022944,9026783,PhysRevLett.113.080401,PhysRevA.95.022335,zhang2,PhysRevA.96.062334}, quantum process tomography \cite{WANG2019269,PhysRevA.63.020101,xiao} and quantum control \cite{DONG2022}.

When  the operators describing a detector are diagonal in the Fock state basis, they are called phase-insensitive (otherwise phase-sensitive) detectors and can be straightforwardly identified using function fitting \cite{Renema:12} or convex optimization \cite{Feito_2009,Lundeen2009,natarajan2013quantum}. For phase-sensitive detectors, generally they can not be simultaneously diagonalized and their reconstruction is thus more complicated. Existing methods include Maximum Likelihood Estimation \cite{PhysRevA.64.024102,PhysRevLett.93.250407}, linear regression \cite{Grandi_2017}, convex-quadratic optimization  \cite{zhang2012mapping,Zhang_2012}, and analytical two-stage solution \cite{wang2019twostage}. Specially, binary detectors can always be simultaneously diagonalized and thus their estimation has an analytical scheme based on Frobenius-norm projection \cite{9029759}.

For $ d $-dimensional QDT, we prepare $ M $ different types of quantum states and the total number of copies of these states  $ N $  is called resource number. Many identification
algorithms assume the experimental resource is diverse enough in QDT, i.e., any detector can be uniquely determined by the measurement outcome statistics. This scenario is called informationally complete (I.C.) \cite{ic1,ic2} and the opposite scenario is called information incomplete (I.I.). In practice, the I.C. condition may not be satisfied  for QDT, which  results in an I.I. scenario (e.g. when $ M<d^2 $ for a $ d $-dimensional detector). In the I.I. scenario and in certain I.C. scenarios where the probe states lie close to the I.I. set although they are still in I.C. set, the QDT problem is ill-conditioned.
To solve this problem,   convex optimization methods with regularization were proposed in \cite{Feito_2009,Lundeen2009} for phase-insensitive detectors and in \cite{zhang2012mapping,Zhang_2012} for phase-sensitive detectors. In experiments, a regularized
least-square method was  used in \cite{add1,add2} for phase-insensitive detectors. However, there is still a lack of closed form solutions for QDT with regularization in these existing methods. To solve this problem, we develop QDT with regularization inspired by  classical transfer function identification. In the previous literature,
a kernel-based regularization was proposed in \cite{PILLONETTO201081,PILLONETTO2011291,CHEN20121525,Pillonetto2014,mazzoleni2022kernel}, which can  cope with bias–variance trade-off. For kernel-based regularization, an important problem is how to design a suitable kernel matrix. Refs. \cite{CHEN20121525,Chen2018} proposed different kernels and Refs. \cite{Chen2013,Mu2018,CHEN2021109682,mu2021asymptotic} discussed how to tune hyper-parameters  in the kernel matrix and the asymptotic properties of these parameters. Further work about kernel-based regularization was studied in  \cite{MU2018327,Pillonetto2016,chentac,pillonetto2022regularized}.

In this paper, we develop regularization methods in QDT which are applicable to both phase-insensitive and phase-sensitive detectors. We  give a closed form solution, applicable to both the cases of I.C. and I.I.. 
We then discuss different regularization forms and explain the advantages of using regularization in QDT.   We consider no regularization as a special case.
In the I.C. scenario, a common step (see e.g. \cite{wang2019twostage,9029759}) is to uniformly distribute the resource  for each quantum state as $ N/M $, which is often not the optimal distribution. Without regularization, we discuss how to optimize the resource  distribution for different types of probe states based on minimizing the mean squared error (MSE) of QDT. We convert this optimization problem into a semidefinite programming (SDP) problem,  which can be solved efficiently. In comparison, if the resource  distribution is given, the probe state design problem was discussed in \cite{xiao2021optimal}. In both the I.C. and I.I. scenarios,
we also prove that under the static assumption (specific definitions in Section \ref{sec51}), the MSE scales as $ O(1/{N}) $ or tends to a constant, and we characterize when the MSE can reach the optimal scaling $ O(1/{N}) $.
We propose an exact characterization of the best regularization for identifiable parameters to  achieve the minimum MSE, allowing the probe states to be I.C. or I.I.. In the I.C. scenario, we obtain the same  best regularization form as proposed in \cite{CHEN20121525}.
We also prove the best regularization can reach the optimal scaling $ O(1/{N}) $ even in the I.I. scenario. 
Numerical examples demonstrate that the optimization of resource  distribution 
and  regularization can reduce the MSE. Then we give the reason why adaptive rank-1 regularization motivated from the best regularization fails to show an $ O(1/{N})$ scaling in QDT, and we find an indication that full-rank regularization might be better. Finally, we apply our algorithm to quantum optical experiments using two-mode coherent states for binary detectors. The experimental results show that the adaptive regularization has a lower MSE compared with the Tikhonov regularization method in \cite{wang2019twostage}. The main contributions
of this paper are summarized as follows.
\begin{enumerate}
	\item[(i)] A closed form of regularized QDT solution is established with different regularization forms in the I.C. and I.I. scenarios. The motivations and advantages to apply regularization in QDT are discussed.
	\item[(ii)] Without regularization,  we optimize the resource (probe state) distribution by converting it to a semidefinite programming (SDP) problem in the I.C. scenario.
	\item[(iii)] Under the static assumption, we prove that the MSE scales as $ O(1/{N}) $ or tends to a constant and we  characterize when the MSE can reach the optimal scaling $ O(1/{N}) $. In addition, an exact characterization of the best regularization for identifiable parameters to  achieve the minimum MSE is given in the I.C. and I.I. scenarios.
	\item[(iv)] Simulation results are provided to verify the effectiveness of resource distribution optimization and regularized QDT. Quantum optics experimental results are presented to demonstrate the necessity of choosing a proper regularization form to further reduce the QDT error.
\end{enumerate}

This paper is organized as follows. In Section \ref{sec2}, we introduce the background knowledge and weighted least squares for QDT. In Section \ref{sec3}, we discuss different regularization forms for QDT. In Section \ref{sec5}, we characterize the scaling of MSE under static assumptions and the best regularization for identifiable parameters. In Section \ref{ns}, we give  numerical examples and in Section \ref{secexp}, we present experimental results. Conclusions are presented in Section \ref{con}.

Notation: For a matrix $ A $, $A \geq 0$ means $A$ is positive semidefinite. The conjugation and transpose $(T)$
of $A$ is $A^{\dagger}$. The trace
of $A$ is $\operatorname{Tr}(A)$. The rank of $ A $ is  $  \operatorname{Rank}(A) $. The identity matrix is $I$. The real and complex domains are $\mathbb{R}$ and $\mathbb{C}$, respectively. The
tensor product is $\otimes$. The set of all $d$-dimensional
complex/real vectors is $ \mathbb{C}^{d}/\mathbb{R}^{d} $. Row and column vectors also denoted as $  \langle\psi| $ and $|\psi\rangle$, respectively. The Frobenius norm for a matrix and 2-norm for a vector are $\|\cdot\|$.  The Kronecker delta function is $\delta$. $\mathrm{i}=\sqrt{-1}$. The  diagonal matrix $ X $ formed from vector $ b $ is denoted as $ X=\operatorname{diag}(b) $.
For any $X_{d \times d}\geq 0$ with spectral decomposition $X=U P U^{\dagger},$ define $\sqrt{X}$ or $X^{\frac{1}{2}}$ as $U \operatorname{diag}\left(\sqrt{P_{11}}, \sqrt{P_{22}}, \ldots, \sqrt{P_{d d}}\right) U^{\dagger}$. The Pauli matrices are $ \sigma_x, \sigma_y, \sigma_z $.

\section{Preliminaries and identification algorithm}\label{sec2}
Here we present the background knowledge and  briefly introduce the QDT identification  algorithm in \cite{wang2019twostage}. Based on this QDT identification  algorithm, we introduce weighted least squares (WLS) in QDT and explain its advantages.
\subsection{Quantum state and measurement}
For a $d$-dimensional quantum system, its state can be described by a $d \times d$ Hermitian
matrix $\rho$, which satisfies $\rho\geq0$ and $\operatorname{Tr}(\rho)=1$. When $\rho=|\psi\rangle\langle\psi|$ for some $|\psi\rangle\in\mathbb{C}^{d}$, we call $ \rho $ a pure state. Otherwise, $\rho$ is called a
mixed state, and can be represented using pure states $\left\{\left|\psi_{i}\right\rangle\right\}$ as $\rho=\sum_{i} c_{i}\left|\psi_{i}\right\rangle\left\langle\psi_{i}\right|$ where $c_{i} \in \mathbb{R}$ and
$\sum_{i} c_{i}=1 $ with $c_{i}\geq 0  $.

A set of operators $\left\{P_{i}\right\}_{i=1}^{n}$ named Positive-operator-valued measure (POVM)  characterizes a corresponding detector as a measurement device. Each POVM element $P_{i}$
is Hermitian and positive semidefinite, and together they  satisfy the completeness
constraint $\sum_{i=1}^{n} P_{i}=I$. 
When the measurements corresponding to  $\left\{P_{i}\right\}$ are performed on $\rho$, the probability of obtaining the $i$-th result is given by the Born's rule
\begin{equation}
p_{i}=\operatorname{Tr}\left(P_{i} \rho\right).
\end{equation}
From the completeness constraint, we have $\sum_{i} p_{i}=1$. 

\subsection{Problem formulation of QDT}
Suppose the true values of the POVM elements
are $\left\{P_{i}\right\}_{i=1}^{n}$. We design $ M $ different types of quantum states $\rho_{j}$ (called probe states) and record the measurement results $\hat{p}_{i j}$
as the estimate of $p_{i j}=\operatorname{Tr}\left(P_{i} \rho_{j}\right)$. Each probe state has resource number $N_j$ (i.e., $ N_{j} $ copies) and the total resource number is $N=\sum_{j=1}^{M} N_{j} $.  	Given experimental data $\left\{\hat{p}_{i j}\right\}$, the problem of QDT \cite{wang2019twostage} can be formulated as 
\begin{equation}
\min _{\left\{\hat{P}_{i}\right\}_{i=1}^n} \sum_{i=1}^{n} \sum_{j=1}^{M}\left[\hat{p}_{i j}-\operatorname{Tr}\left(\hat{P}_{i} \rho_{j}\right)\right]^{2}
\end{equation}
such
that $\hat P_i=\hat P_i^{\dagger},\hat{P}_{i} \geq 0 $  for  $1 \leq i \leq n$ and $\sum_{i=1}^{n} \hat{P}_{i}=I$.

Let $\left\{\Omega_{i}\right\}_{i=1}^{d^{2}}$ be a complete
basis set of orthonormal operators with dimension $d$. informationally let $	\operatorname{Tr}\left(\Omega_{i}^{\dagger} \Omega_{j}\right)=\delta_{i j},\Omega_{i}=\Omega_{i}^{\dagger}  $
and $\operatorname{Tr}\left(\Omega_{i}\right)=0$ except $ \Omega_{1}=I / \sqrt{d} $. Then we can parameterize the detector
and probe states as
\begin{equation}
\label{para}
\begin{aligned}
P_{i} =\sum_{a=1}^{d^{2}} \lambda_{i}^{a} \Omega_{a},\rho_{j} =\sum_{b=1}^{d^{2}} \phi_{j}^{b} \Omega_{b}.
\end{aligned}
\end{equation}
Using Born's rule, we can obtain
\begin{equation}
p_{i j}=\sum_{a=1}^{d^{2}} \phi_{j}^{a} \lambda_{i}^{a} \triangleq \phi_{j}^{T} \lambda_{i},
\end{equation}
where $\phi_{j}\triangleq\left(\phi_{j}^{1}, \ldots \phi_{j}^{d^{2}}\right)^{T}$ and $\lambda_{i}\triangleq\left(\lambda_{i}^{1}, \ldots \lambda_{i}^{d^{2}}\right)^{T}$  are the parameterization vectors of  $ \rho_{j} $ and   $ P_i $, respectively.
Suppose the outcome for $P_{i}$ of $ \rho_j $ appears $n_{i j}$ times, and then $\hat{p}_{i j}=n_{i j} /N_{j} $. Denote
the estimation error as $e_{i j}=\hat{p}_{i j}-p_{i j} $. According to the central limit theorem, $ e_{ij} $ converges in distribution
to a normal distribution with mean zero and variance $ \left(p_{i j}-p_{i j}^{2}\right) /N_j$. We thus have the least squares (LS) equation 
\begin{equation}
\hat{p}_{i j}=\phi_{j}^{T} \lambda_{i}+e_{i j}.
\end{equation}

To propose least squares (LS) and weighted least squares (WLS) solutions in QDT,  in the following we  write down and solve the linear equation for each POVM element \emph{individually}. This can be achieved by rearranging the data after implementing all the measurements. Collect the parameterization of the probe states as $X=\left(\phi_{1}, \phi_{2}, \ldots, \phi_{M}\right)^{T}$. 
For the $ i $-th POVM element $ P_i $, let 
\begin{equation*}
\begin{aligned}
\hat{y}_{i}&\triangleq\left(\hat{p}_{i 1}, \hat{p}_{i 2}, \ldots, \hat{p}_{i M}\right)^{T},\\
y_{0}&\triangleq\left((1, \ldots, 1)_{1 \times M}\right)^{T}=\sum_{i} \hat{y}_{i},\\
\mathcal d_{d^{2} \times 1}&\triangleq(\sqrt{d}, 0, \ldots, 0)^{T},\\
e_{i}&\triangleq\left[e_{i 1}, \ldots, e_{i M}\right]^{T}.
\end{aligned}
\end{equation*}
Define $ {\bar { y_i}}\triangleq\hat{ y}_{i}-\frac{1}{n} y_{0}$ and $ \theta_{i}\triangleq \lambda_i-\frac{1}{n}{\mathcal d}$. Then we have
\begin{equation}
\label{eqsmall}
\bar{y_i}=X\theta_{i}+e_i.
\end{equation} 
Now the QDT problem can be transformed into the following form:
\begin{problem}\label{problem2}
	For $ 1\leq i\leq n $,	given experimental data $ \bar{y}_{i} $, solve $ \min _{\hat{P}_{i}}\|\bar{y}_{i}-X\theta_{i}\|^{2}  $  with $\hat{P}_{i} \geq 0 $, where $ \lambda_i= \theta_{i}+\frac{1}{n}{\mathcal d} $ is the parametrization of $ \hat{P}_{i} $. 
\end{problem}

\subsection{Weighted least squares in QDT}
To solve Problem \ref{problem2}, the standard LS solution is
\begin{equation}\label{ls1}
\hat \theta_{i,\text{LS}}=\left(X^{T} X\right)^{-1} X^{T}\bar{y_i},
\end{equation}
and then the estimator for each detector is $\hat \lambda_{i,\text{LS}}=\hat \theta_{i,\text{LS}}+ \frac{1}{n}{\mathcal d} $, which is equivalent to equation (9) in \cite{wang2019twostage}. 

Although all the estimation errors $ e_{ij} $ have  zero mean, they have  different variances, which is called heteroscedasticity in statistics. The constrained least squares as equation (6) in \cite{wang2019twostage} and standard LS  \eqref{ls1} do not consider heteroscedasticity. However, WLS consider the heteroscedasticity property and has optimal MSE. We thus consider WLS estimate
\begin{equation}
\label{wls}
\hat{\theta}_{i,\text{WLS}}=\left({X}^{T}W_i {X}\right)^{-1} {X}^{T}W_i \bar y_i,
\end{equation}
where
\begin{equation}
{W_i}= \operatorname{diag}\left(\left[\frac{N_{1}}{p_{i1}-p_{i1}^{2}}, \ldots, \frac{N_{M}}{p_{iM}-p_{iM}^{2}}\right]\right) 
\end{equation}
is the weighting matrix. We assume that $ p_{ij} $ is not equal to $ 0 $ or $ 1 $, which is reasonable because  $ p_{ij} \in [0,1] $ and generally the probability for $ p_{ij}=0 $ or $ 1 $ is $ 0 $ in theory. The following are the two main advantages of using WLS:
\begin{itemize}
	\item We can normalize the estimation errors to normal Gaussian errors and solve the heteroscedasticity problem.
	With increasing measurements, each $ e_{ij} $ will converge asymptotically  to a Gaussian random
	variable with mean zero and variance $\sigma_{ij}=\left(p_{ij}-p_{ij}^{2}\right) / N_{j}$. Thus, we have
	$\mathbb{E}\left({e_{i}e_{i}}^{T}\right)={W_i}^{-1}$. Define $ Q_{i}\triangleq\sqrt{W_{i}}^{-1}/\sigma $ for certain $ \sigma>0 $.
	Then we multiply by $ Q_{i}^{-1} $ in \eqref{eqsmall} as
	\begin{equation}
	Q_{i}^{-1}{\bar y_i}=Q_{i}^{-1}{X} {\theta_i}+Q_{i}^{-1}{e_i}.
	\end{equation}
	Let $ \mathbb{E}(\cdot) $ denote the expectation with respect
	to all possible
	measurement results. The new errors have an independent identical   normal distribution (i.i.d.) with
	\begin{equation}\label{sigma2}
	\mathbb{E}\left({Q_{i}}^{-1}{e_{i}e_{i}}^{T} {Q_{i}}^{-1}\right)=\sigma^{2}I.
	\end{equation}
	Thus,  all the variances of the estimation errors are normalized to $ \sigma^2 $.

	\item For any unbiased linear estimator $ \hat \theta_i  $ for $ \theta_i $, we have \cite{MU2020108837}
	\begin{equation}
	\begin{aligned}
	\operatorname{MSEM}\left(\hat{\theta}_{{i,\text{WLS}}}\right)&=\mathbb{E}\left[\left(\hat{\theta}_{{i,\text{WLS}}}-\theta_i\right)\left(\hat{\theta}_{i,\text{WLS}}-\theta_i\right)^{T}\right]\\
	&=\left(X^{T}W_iX\right)^{-1}\leqslant  \operatorname{MSEM}\left(\hat{\theta}_{{i}}\right),
	\end{aligned}
	\end{equation}
	where $ \operatorname{MSEM}\left(\cdot\right) $ is the MSE matrix.
	The MSE of all the POVM elements is 
	\begin{equation}
	\label{upmse}
	\begin{aligned}
	\mathbb{E}\left(\sum_{i=1}^{n}\left\|\hat{P}_{i}-P_{i}\right\|^{2}\right) 
	&=\sum_{i=1}^{n}\mathbb{E}\left(\left\|\hat{\theta}_i-\theta_i\right\|^{2}\right) \\
	&=\sum_{i=1}^{n}\operatorname{Tr}\left(\operatorname{MSEM}\left(\hat{\theta}_{i}\right)\right).
	\end{aligned}
	\end{equation}
	Hence, the WLS solution to Problem \ref{problem2} has the minimum MSE.
\end{itemize}

In practice, the weighting matrix $W_i$  is unknown and a feasible
solution is to use the asymptotic estimate
\begin{equation}\label{hatw}
\hat{W}_i=\operatorname{diag}\left(\left[\frac{N_{1}}{\hat{p}_{i1}-\hat{p}_{i1}^{2}}, \ldots, \frac{N_{M}}{\hat{p}_{iM}-\hat{p}_{iM}^{2}}\right]\right).
\end{equation}
Denote $\hat Q_{i}^{-1}\triangleq\sqrt{\hat W_{i}}^{-1}/\sigma  $, $ \tilde{y_{i}}\triangleq\hat Q_{i}^{-1}{\bar y_i}$, $\tilde{X}_{i}\triangleq \hat Q_{i}^{-1}{X}$, $\tilde{e_i}\triangleq \hat Q_{i}^{-1}{e_i}  $
and the  model  equivalent to \eqref{eqsmall} is
\begin{equation}
\label{weightmodel}
{\tilde y_i}={\tilde X}_i {\theta_i}+\tilde{e}_{i},
\end{equation}
where the variance of $ \tilde{e}_{i} $ is $  \sigma^{2}I$
and the practical asymptotic WLS (AWLS) estimate  is 
\begin{equation}
\label{awls}
\begin{aligned}
\hat{\theta}_{i,\text{AWLS}}&=\left({X}^{T}\hat W_i {X}\right)^{-1} {X}^{T}\hat W_i \bar y_i\\
&=\left(\tilde X_{i}^{T} \tilde X_{i}\right)^{-1}\tilde X_{i}^{T}\tilde{y_i}.	
\end{aligned}
\end{equation} 
The difference between $\hat{{\theta}}_{{i,\text{AWLS}}}$ and $\hat{{\theta}}_{{i,\text{WLS}}}$ is asymptotically small in comparison with $ \hat{{\theta}}_{{i,\text{WLS}}}$  \cite{MU2020108837}. Thus, the estimate \eqref{awls} is accurate enough and asymptotically
coincides with \eqref{wls}. Using the LS estimate \eqref{ls1} or WLS estimate \eqref{awls}, we can obtain a POVM estimate $ \hat E_i=\sum_{a=1}^{d^2}\left(\hat{\theta}_{i, \text{LS/WLS}}+\frac{1}{n}{\mathcal d}\right)_{a} \Omega_a $ and
	\begin{equation}
	\mathbb{E}\|\hat E_i- P_i\|^2=\operatorname{Tr}\left(\operatorname{MSEM}\left(\hat{\theta}_{i}\right)\right).
	\end{equation}
	We call the error $ \mathbb{E}\|\hat E_i- P_i\|^2 $  the \emph{LS MSE} for the $ i $-th POVM element.
	Note that $ \{\hat E_i\}_{i=1}^{n} $ may have negative eigenvalues due to the noise or error in the measurement results. Thus, we need further correction to obtain a positive semidefinite  estimate $ \{\hat P_i\}_{i=1}^{n} $ and  in this paper, we utilize the algorithm in \cite{wang2019twostage} to achieve this.
	We refer to the error $ \mathbb{E}\|\hat P_i- P_i\|^2 $ as the \emph{final MSE} for the $ i $-th POVM element.

\begin{rem}
	One may notice that \eqref{eqsmall} has the same linear
	regression form $ y=X\theta+e $ as transfer function identification in system identification \cite{CHEN20121525}. 
	However, there are some differences between QDT and transfer function identification for classical (non-quantum) systems. First, in QDT, more measurement data will only enhance the data accuracy in $ y $ and the dimension of $ y $ is fixed with given probe states. In transfer function identification,  the dimension of $ y $ increases for more data. Second, the  parameterization matrix $ X $  is determined by the given probe states and $ X^{T}X $ can be singular (e.g., $ M<d^2 $) in QDT. In transfer function identification, $ X $ depends on the input data and measurement data.  In practice, $  X^{T}X  $ is therefore always invertible but the condition number may be  large. Thus, the standard LS cannot give an accurate estimate. Finally, the variance of the noise $ e $ is often assumed to be a constant in transfer function identification. However, in QDT, the variances of noise are usually different and decrease as $ O(1/{N}) $  where $ N $ is the resource number.
\end{rem}

\section{Regularization in QDT}\label{sec3}
In QDT, when  the different types of probe states are similar or I.I., leading to an ill-conditioned problem,
convex optimization methods with regularization were proposed in \cite{Feito_2009,Lundeen2009} for phase-insensitive detectors and in \cite{zhang2012mapping,Zhang_2012} for phase-sensitive ones. The motivation of introducing regularization is to mitigate the ill-conditioned property. For phase-insensitive detectors, the regularization form is chosen such that the diagonal elements of the reconstructed detector have smooth variations \cite{Zhang_2012}.
However, for phase-sensitive detectors, a suitable regularization form is not easy to find. In addition, convex optimization methods cannot give a closed form solution. Therefore, in this section, we use regularization in the WLS of QDT which can give a closed form solution.
\subsection{Regularized weighted least squares}
In the ill-conditioned scenario, the condition number of $ \tilde X_{i}^{T}\tilde X_{i}$ can be large or even infinite.
To solve this problem, we add regularization in the weighted model \eqref{weightmodel} as
\begin{equation}
\left\|\tilde y_{i}-\tilde X_{i} \theta_{i}\right\|^{2}+\theta_{i}^{T} D_{i} \theta_{i},
\end{equation}
where $D_{i}$ is positive semi-definite and called a regularization
matrix.   
Denote $ R_i\triangleq\tilde X_{i}^{T}\tilde X_{i} $. After we add regularization,  the estimate is changed to be
\begin{equation}
\begin{aligned}
\hat{\theta}_{i,\text{RWLS}}&=\left(R_i+D_{i}\right)^{-1} \tilde X_{i}^{T}\tilde y_i.\\	
\end{aligned}
\end{equation}
The expectation of  $ \hat{\theta}_{i,\text{RWLS}} $ is 
\begin{equation}
\mathbb{E} \left(\hat{\theta}_{i,\text{RWLS}}\right)=\left(R_{i}+D_{i}\right)^{-1} R_{i} {\theta}_{i}.
\end{equation}
The bias is
\begin{equation}\label{bias}
\theta_{i,\text{RWLS}}^{\text{bias}}\triangleq\mathbb{E} \left(\hat{\theta}_{i,\text{RWLS}}\right)-\theta_{i}=-\left(R_{i}+D_{i}\right)^{-1} D_{i} \theta_{i}.
\end{equation}
Define
\begin{equation}
\begin{aligned}
\tilde{\theta}_i\triangleq&\hat{\theta}_{i,\text{RWLS}}-\mathbb{E} \left(\hat{\theta}_{i,\text{RWLS}}\right)\\
=&\left(R_{i}+D_{i}\right)^{-1}\tilde{X}_{i}^{T}\left(\tilde{y}_{i}-\tilde{X}_{i} \theta_{i}\right)\\
=&\left(R_{i}+D_{i}\right)^{-1}\tilde{X}_{i}^{T}\tilde{e}_{i},
\end{aligned}
\end{equation}
and then the MSE matrix of  $ \hat{\theta}_{i,\text{RWLS}} $ is
\begin{equation}\label{msed}
\begin{aligned}
&\operatorname{MSEM}\left(\hat{\theta}_{i,\text{RWLS}}\right)=\mathbb{E}\left[\left(\hat{\theta}_{i,\text{RWLS}}-\theta_{i}\right)\left(\hat{\theta}_{i,\text{RWLS}}-\theta_{i}\right)^{T}\right] \\
&=\mathbb{E} \left(\tilde{\theta}_i \tilde{\theta}_{i}^{T}\right)+\theta_{i,\text{RWLS}}^{\text{bias}}\left(\theta_{i,\text{RWLS}}^{\text{bias}}\right)^{T}\\
&=\left(R_i+D_i\right)^{-1}\left(\sigma^{2} R_i+D_i \theta_{i} \theta_{i}^{T} D_{i}^{T}\right)\left(R_i+D_i\right)^{-1}.
\end{aligned}
\end{equation}
An   MSE matrix similar to \eqref{msed} can be found in \cite{CHEN20121525} for transfer function identification with standard LS estimation.  The LS MSE of QDT is $ \operatorname{Tr}(\text{MSEM}) $ and  depends on the true parameter $ \theta_{i} $. 
When the probe states are I.C., we can obtain an estimate without regularization (i.e., $ D_i=0 $) and the MSE matrix becomes 
\begin{equation}\label{mseic}
\begin{aligned}
&\operatorname{MSEM}\left(\hat{\theta}_{i,\text{RWLS}}\right)=\sigma^{2}R_i^{-1},
\end{aligned}
\end{equation}
which is independent of the true parameter $ \theta_{{i}} $.

	 Based on the development in classical system identification, several motivations of  applying regularization in QDT  are as follows:
	\begin{enumerate}
	\item[(i)] Regularization is a typical solution to ill-conditioned problems. In the field of classical transfer function identification (see e.g., \cite{Chen2013}), the input signal is band-limited, and then the matrix $R_i$ may become ill-conditioned as the amount of data increases. Similarly in QDT, the input probe states can be ``band-limited", in the sense that the types of the probe states are not rich enough (especially when coherent states are employed) which leads to the conversion from I.C. to I.I. This current incapability of realizing perfect number states endows $R_i$ with a large condition number, which can be reduced by regularization while still maintaining a closed form solution.
	\item[(ii)] From an alternative point of view, regularization leverages the bias-variance trade-off. The regularization estimation is biased as \eqref{bias}, which can lead to an MSE smaller than that of the standard LS estimation both in the I.C. and I.I. scenarios.
\end{enumerate}

	There are  also differences of applying regularization between  QDT and classical system identification. All physical POVM elements must be positive semidefinite and sum to identify, which may affect or even guide the design of the specific regularization form in QDT. For example, Ref. \cite{wang2019twostage} noted that POVM elements satisfying these physical constraints always have eigenvalues in $[0,1]$. Direct LS estimation for ill-conditioned QDT usually gives a large $ \|\hat{\theta}_i\| $, which may have eigenvalues outside $[0,1]$ and become nonphysical. Therefore, the regularization $ \theta_{i}^{T} D_{i} \theta_{i} $ is added to the cost function in \cite{wang2019twostage} as a penalty term, promoting the satisfaction of the physical constraints. Apart from this, other differences will be detailed in Sec. \ref{difference}.

Regularized weighted regression is also applied in quantum state tomography. For example, in \cite{MU2020108837}, 
	their motivation is that the quantum state $ \rho $ is usually of low rank \cite{addre1} and thus it is reasonable to add a Tikhonov regularization as Sec. \ref{tik}. However, in QDT, the POVM elements are  not always of low rank. For example, in the continuous-variable optical experiment in the paper and in \cite{Feito_2009,Lundeen2009,Zhang_2012}, the POVM elements are all full-rank.
	Thus we introduce and discuss  more regularization forms  in Sec. \ref{dre}.

\subsection{Different regularization forms in QDT}\label{dre}
Here we discuss different regularization forms in QDT. Firstly, we consider no regularization (i.e., $ D_i=0 $) as
a special regularization form in the I.C. scenario. Since the MSE in \eqref{mseic} does not depend on true parameter $ \theta_{{i}} $, we propose resource distribution optimization of $ N_j $ to minimize the LS MSE with given probe states. Then we present some common regularization forms. With regularization, the LS MSE in \eqref{msed} depends on true parameter $ \theta_{{i}} $. Thus we cannot optimize resource distribution as without regularization and we use a uniformly distributed $ N_j=N/M  $.
\subsubsection{Without regularization}
Without regularization, Refs. \cite{wang2019twostage,9029759}  choose $ N_j=N/M $ for given probe states, which is often not the optimal distribution.  Similar input design problems in classical systems and control have been widely studied and there are many existing results, e.g., D,A,E-optimal input design \cite{Boyd2004Convex}. Here, we formulate and solve the problem within the framework of  A-optimal design problem, where the trace of the covariance matrix (i.e., MSE) is minimized.

Let $ \eta_j=\frac{N_{j}}{N} $, and the optimization of resource distribution problem can be formulated as
\begin{equation}
\label{rs}
\begin{aligned}
\min_{\{\eta_{j}\}_{j=1}^{M}} & \sum_{i=1}^{n}\operatorname{Tr}\left( \sum_{j=1}^{M}\left(\eta_{j}w_{ij} \phi_{j} \phi_{j}^{T}\right)\right)^{-1} \\
\text { s.t. } & \eta_{j} \geq 0, \sum_{j=1}^{M} \eta_{j}=1,
\end{aligned}
\end{equation}
where $ \phi_{j} $ is  the given parameterization vectors of  $ \rho_{j} $ and $ w_{ij}  $ is the weighted constant which we may obtain from a prior information. If we do not have a prior information, we can set $ w_{ij}=1 $.
This optimization problem  is  convex  and it can be converted to a semidefinite programming (SDP) problem
\begin{equation}
\label{rs2}
\begin{aligned}
&\min_{\{\eta_{j}\}_{j=1}^{M},\{u_{k}\}_{k=1}^{d^2}} \sum_{k=1}^{d^2} u_k \\
&\quad\quad\text { s.t. }  \left[\begin{array}{cc}
\sum_{j=1}^{M}\eta_{j} w_{ij}\phi_{j} \phi_{j}^{T} & v_{k} \\
v_{k}^{T} & u_{k}
\end{array}\right] \geq 0,\\
&\quad\quad 1 \leq k \leq d^{2}, 1 \leq i \leq n, \\
&\quad\quad\eta_{j} \geq 0, \sum_{j=1}^{M} \eta_{j}=1,
\end{aligned}
\end{equation}
where $v_{k}$ is the $k$-th unit vector.
Using CVX \cite{cvx,gb08}, we can solve \eqref{rs2} efficiently. Note that $ N_j=\eta_jN $ may not be an integer, and we need to round it up or down. In comparison, if the resource  distribution is given, the probe state design problem was discussed in \cite{xiao2021optimal} based on minimizing an upper bound on the MSE and the condition number. 
\subsubsection{Tikhonov regularization}\label{tik}
A most common regularization form is in a Tikhonov
sense \cite{Boyd2004Convex}. In QDT, a natural method is to choose regularization matrix as 
\begin{equation}	\label{Ti}
D_i^{\text{Tikhonov}}=cI,
\end{equation}
where $ c $ is a positive constant.  Ref. \cite{wang2019twostage} did not use WLS and chose $D_i=\frac{c}{N}I  $ which is Tikhonov regularization, because
\begin{equation}\label{wangtik}
\begin{aligned}
\hat{\theta}_{i,\text{RWLS}}&=\left(X^{T}X+\frac{c}{N}I\right)^{-1} X^{T} \bar y_i\\
&=\left(X^{T}NI X+{c}I\right)^{-1}  X^{T} NI\bar y_i,\\	
\end{aligned}
\end{equation}
where the weighted matrix is $ NI $ instead of \eqref{hatw}.

\subsubsection{Kernel-based regularization}\label{difference}
In transfer function identification, Refs. \cite{PILLONETTO201081,PILLONETTO2011291,CHEN20121525,Pillonetto2014} proposed kernel-based regularization and explained regularization in a Bayesian perspective. We assume the true parameter $ \theta_{i}$ is a random variable and has a Gaussian distribution with zero mean and
covariance matrix $ S_i $:
\begin{equation}
\theta_{i} \sim \mathcal{N}\left(0, S_i\right).
\end{equation}
Therefore, the posterior estimate is 
\begin{equation}\label{rwlssolution}
\begin{aligned}
\hat{\theta}_{i}^{\text {post }}=&\left(S_iR_{i}+\sigma^{2} I\right)^{-1} S_i F_{i} \\
=&\left(R_{i}+\sigma^{2} S_{i}^{-1}\right)^{-1} F_{i},
\end{aligned}	
\end{equation}
where $ F_{i}\triangleq\tilde{X}_{i}^{T}\tilde{y_i} $. If $ S_i $ is singular,  we can use the first equality of \eqref{rwlssolution} to obtain the estimate.
This posterior estimate  is the same as the regularized estimate 
if the regularization matrix $D_i$ is chosen as \cite{CHEN20121525}
\begin{equation}
D_i=\sigma^{2} S_{i}^{-1}.
\end{equation}
This gives an insight into how to choose the regularization matrix $ D_i $ or kernel matrix $ S_{i} $: Let it reflect the  correlations of the parameters \cite{CHEN20121525}.

To use the kernel-based regularization in QDT, we need to solve two problems
\begin{enumerate}
	\item[(i)] In the Bayesian perspective for kernel-based regularization, the mean of the unknown parameters is zero. But in QDT, the mean of the unknown parameters $ \lambda_i $ is usually  not zero.
	\item[(ii)] Heteroscedasticity:  In transfer function identification, it is usually assumed that the noises have the same variances. But the estimation errors $ e_{ij} $ usually have different variances in QDT.
\end{enumerate}
The first problem is solved by modeling in \eqref{eqsmall} where the unknown parameter $ \theta_i $
becomes zero-mean. For the second problem, WLS \eqref{weightmodel} solves the heteroscedasticity problem.

There are two advantages of using kernel-based regularization in QDT compared with using kernel-based regularization  in transfer function identification:
\begin{enumerate}
	\item[(i)]  In transfer function identification, we need to identify the variance of the noise firstly, while we already know the approximate variance of the estimation error in QDT from measurement data. 
	\item[(ii)]  In transfer function identification, the problem dimension increases as more data are generated, resulting in increased difficulty. While in QDT, more data will only enhance the data accuracy and the dimension is fixed with given probe states. 
\end{enumerate}

One limit using kernel-based regularization in QDT is that without prior knowledge the parameter $ \theta_i$ does not have the property of impulse responses of transfer functions  which usually decay exponentially \cite{CHEN20121525}. In this paper, we mainly choose DI kernel which only represents the auto-correlation for each coefficient of QDT as
\begin{equation}
\label{DI}
S_i^{\text{DI}}(k, j)=\left\{\begin{array}{ll}
c \mu^{k}, & \text {if }  k=j,\\
0, & \text {otherwise},
\end{array}\right.
\end{equation}
where $ c \geq 0$, $0 \leq \mu \leq 1 $.
If we have more prior knowledge such as the correlation between different coefficients, we can design more suitable kernels as in transfer function identification. For example, when the detector is close to a phase-insensitive detector, i.e., the POVM elements are close to diagonal matrices in the Fock state basis, the true value $ \theta_{{i}} $ is close to sparse,  which is similar to the decay behavior of impulse responses for stable transfer functions in system identification. Therefore, we can apply TC and  DC kernels \cite{CHEN20121525,Chen2013} in
	transfer function identification
	\begin{equation}\label{tc}
	S_i^{\text{TC}}(k, j)=c \min \left(\mu^j, \mu^k\right),
	\end{equation}
	where $ c \geq 0$, $0 \leq \mu \leq 1 $ and 
	\begin{equation}\label{dc}
	S_i^{\text{DC}}(k, j)=c \mu_{1}^{|k-j|} \mu_{2}^{(k+j) / 2},
	\end{equation}
	where $ c \geq 0$, $-1 \leq \mu_1 \leq 1 $ and $0 \leq \mu_2 \leq 1 $.

\subsubsection{Best regularization (in the I.C. scenario)}
For true parameter $ \theta_i $, two natural questions are whether there exists an optimal regularization matrix  and if there exists an optimal regularization matrix, does it depend on  $ \theta_i $?  Ref. \cite{CHEN20121525} has discussed these problems in transfer function identification and  the result also holds for QDT.
The MSE matrix in \eqref{msed} can be rewritten using $ S_i $  as 
\begin{equation}\label{msesi}
\begin{aligned}
\operatorname{MSEM}\left(\hat{\theta}_{i,\text{RWLS}}\right)=&\left(S_iR_i+\sigma^{2} I\right)^{-1}(\sigma^{2} S_{i}R_iS_{i}\\
&+\sigma^{4} \theta_{i} \theta_{i}^{T})\left(R_iS_{i}+\sigma^{2} I\right)^{-1}.
\end{aligned}
\end{equation}
When $ R_i $ is invertible, the following matrix inequality \cite{CHEN20121525,1658250} 
\begin{equation}\label{ls}
\left.\operatorname{MSEM}\left(\hat{\theta}_{i, \mathrm{RWLS}}\right)\right|_{S_{i}=K} \geq\left.\operatorname{MSEM}\left(\hat{\theta}_{i, \mathrm{RWLS}}\right)\right|_{S_{i}=\theta_{i} \theta_{i}^{T}}
\end{equation}
holds for any $ K \geq 0 $. Later, in Theorem \ref{theorem2}, we will extend this inequality to the case where $ R_i $ is singular.
Thus, ideally the  best choice of regularization always includes
\begin{equation}\label{best}
S_i^{\text{best}}=\theta_{i} \theta_{i}^{T},
\end{equation}
which yields the corresponding optimal regularized estimate
\begin{equation}
\hat{\theta}_{i}^{\text{best}}=\left(\theta_{i} \theta_{i}^{T} R_i+\sigma^{2} I\right)^{-1} \theta_{i} \theta_{i}^{T} F_i,
\end{equation}
with $ R_i=\tilde X_{i}^{T}\tilde X_{i} $ and $ F_{i}=\tilde{X}_{i}^{T}\tilde{y_i} $. The theoretically best regularization depends on the unknown parameter and cannot be used in practice.

A natural question is that, is $ \theta_{i} \theta_{i}^{T} $ the only choice for $ S_{i} $ to result in the best regularization?
Ref. \cite{1658250} has given a positive answer for the I.C. scenario. For the I.I. scenario we will give a negative answer in Sec. \ref{sec52}.
\subsubsection{Adaptive regularization}
As motivated by the best regularization, we  can propose adaptive regularization with rank-1 kernel matrix which is similar to the rank-1 kernel matrix for transfer function identification in \cite{chentac}. Firstly, we consider a two-step adaptive regularization. In the first step, we use Tikhonov or kernel-based regularization and we can obtain a rough estimate $ \hat\theta_{i}^{0} $ with certain kernel matrix $ S_i^{(1)} $. Then in the second step, we repeat using the measurement data in the first step, but now the regularization matrix is adaptively chosen as  
\begin{equation}
\label{first}
S^{\text{rank-1}}_{i}=\hat\theta_{i}^{0}\left(\hat\theta_{i}^{0}\right)^T.	
\end{equation}
The following analysis and Theorem \ref{theorem1} in the next section indicate that full-rank kernel matrix may be better than rank-1 kernel matrix, because a full-rank $ S_i $ does not induce a dimension reduction from $ R(B) $ to $ R(S_iB) $. Therefore, we also consider to use full-rank kernel matrix as
\begin{equation}\label{full}
S_{i}^{\text{full-rank}}=S^{\text{rank-1}}_{i}+S^{\text{DI/TC/DC}}_{i},
\end{equation}
in Sec. \ref{ns}.

It is an important problem to determine the kernel matrix and some different kernels are proposed in transfer function identification. For a structure-given kernel matrix, optimization of the hyper-parameters (such as $c$, $ \mu $ in \eqref{DI}) in the kernel matrix has been discussed in \cite{chentac,Chen2018,Chen2013}. However, the question of how to choose the optimal adaptive kernel matrix is still an open problem. 

\section{Characterizing the MSE of QDT with regularization }\label{sec5}
\subsection{On the MSE scaling}\label{sec51}
To analyze the performance of different regularization methods, we characterize the asymptotic behavior of the estimation error, e.g., MSE. Without loss of generality, we can always normalize the variances of the estimation errors to $ 1 $, i.e., $ \sigma^{2}=1 $ in \eqref{sigma2}. We give the
following assumptions.
\begin{assumption}\label{assum1}
	The probe state parameterization matrix $ X $ is given. The kernel matrix $S_i$ is given. For each $1\leq j\leq n$, $\lim _{N \rightarrow \infty} \frac{N_{j}}{N}= h(j)$ where $ h(j) $ is a constant in $ [0,1] $ depending on $ j $.
\end{assumption}
We refer to Assumption  \ref{assum1} as the \emph{static assumption}. With Assumption \ref{assum1}, the probe state parameterization matrix  and kernel matrix are given as constant matrices which do not change in our analysis and the resource distribution for each probe state can change as $ N $ increases. But the limit of the ratio is a constant and can be $ 0 $ or $ 1 $.
We say that the random sequence $\left\{\xi_{N}\right\} $ converges almost surely to a random variable $\xi $ if $\operatorname{P}\left(\lim_{N \rightarrow \infty}\left\|\xi_{N}-\xi\right\|_{2}=0\right)=1$, which can be written as $\xi_{N} \stackrel{a. s.}{\rightarrow} \xi$ as $N \rightarrow \infty$.
For the weighted matrix $\hat{W}_{i}  $, its deviation from the true value $ {W}_{i} $ has been derived in \cite{MU2020108837} as
\begin{equation}\label{weight}
\begin{aligned}
\hat{W}_{i}=&\operatorname{diag}\left(\left[\frac{N_{1}}{\hat{p}_{i 1}-\hat{p}_{i 1}^{2}}, \ldots, \frac{N_{M}}{\hat{p}_{i M}-\hat{p}_{i M}^{2}}\right]\right)\\
=&\left(1+O\left(\frac{1}{\sqrt{N}}\right)\right)W_{i}.\\
\end{aligned}
\end{equation}
We define 
\begin{equation}
B\triangleq\lim _{N \rightarrow \infty}\frac{X^{T}  W_i X}{N}, \hat B_{N}\triangleq\frac{X^{T} \hat W_i X}{N},
\end{equation}
where the normalized weighted parameterization matrix $\hat B_{N}=\left(1+O\left(1/{\sqrt{N}}\right)\right) B  $ for constant matrix $ B $  because  $\lim _{N \rightarrow \infty} \frac{N_{j}}{N}= \operatorname{constant} $. Therefore, $ \hat B_{N}\stackrel{a. s.}{\rightarrow}B $ as $ N \rightarrow \infty $.

We denote $ R(X) $ as the range space of $ X $ and $ N(X) $ as the null space of $ X $. Then we propose the following theorem to characterize the MSE.
\begin{theorem}\label{theorem1}
	In the regularization-based QDT, if the $i$-th POVM element satisfies the static assumption, then its LS MSE $ \mathbb{E}\left\|\hat{E}_i-P_i\right\|^2 $ and final MSE $ \mathbb{E}\left\|\hat{P}_i-P_i\right\|^2 $ both scale as $ O\left(1/N\right) $ if and only if the true values of the unknown parameters satisfy
	$\theta_i\in R(S_{i} B)  $. Otherwise, the LS MSE $\mathbb{E} \left\|\hat{E}_i-P_i\right\|^2 $ converges to a positive value.
\end{theorem}

\begin{pf}
	For the $ i $-th POVM element, according to \eqref{msesi} and $ \sigma^{2}=1 $, the MSE is
	\begin{equation}
	\begin{aligned}
	&\operatorname{Tr}\left[\operatorname{\left.\operatorname{MSEM}\right|}_{S_{i}}\right]\\
	=&\operatorname{Tr}\big[\left(S_{i} R_{i}+ I\right)^{-1}( S_{i} R_{i} S_{i}+ \theta_{i} \theta_{i}^{T})\left(R_{i} S_{i}+ I\right)^{-1}\big]\\
	=&\operatorname{Tr}\big\{\left[\left(S_{i} R_{i}+ I\right)\left(R_{i} S_{i}+ I\right)\right]^{-1}\left(S_{i} R_{i} S_{i}+ \theta_{i} \theta_{i}^{T}\right)\big\},\\
	\end{aligned}
	\end{equation}
	where $ R_i=X^{T} \hat{W}_{i} X $.
	We define 
	\begin{equation}\label{A1}
	\begin{aligned}
	A_{1}&\triangleq\left(S_{i} R_{i}+ I\right)\left(R_{i} S_{i}+ I\right)\\
	&=\left(NS_{i}\hat{B}_{N}+ I\right)\left(N\hat{B}_{N} S_{i}+ I\right),\\
	\end{aligned}
	\end{equation}
	and
	\begin{equation}\label{A2}
	\begin{aligned}
	A_{2}\triangleq S_{i} R_{i} S_{i}=NS_{i}\hat B_{N}S_{i}.
	\end{aligned}
	\end{equation}
	Now the MSE becomes $ \operatorname{Tr}\left(A_{1}^{-1}(A_{2}+\theta_{i}\theta_{i}^{T})\right) $.

	We then introduce the following lemma
	\begin{lemma}\label{lemma2}
		\cite{WU198853,CUI201717} For an $n \times n$ complex matrix $T$, the following statements are equivalent:
		\begin{enumerate}
			\item $T=A B$, where $A, B \geqslant 0$;
			\item $T=A B$, where $A>0$ and $B \geqslant 0$;
			\item $T$ is similar to a nonnegative diagonal matrix.
		\end{enumerate}
	\end{lemma}
	
	From Lemma \ref{lemma2}, $ S_iB $ is similar to a nonnegative diagonal  matrix and we assume $  S_iB=Q^{-1}\Sigma_{1}Q$ where $\Sigma_{1}=\operatorname{diag}(\Sigma_{11},\Sigma_{12})  $ and $\Sigma_{11}$ is a $ k\times k $ positive  diagonal matrix, $\Sigma_{12}$ is a $ (d^2-k)\times (d^2-k) $ zero matrix. Therefore, $ NS_iB+I $ can also be diagonalized by $ Q $ as
	\begin{equation}\label{inv}
	\begin{aligned}
	N S_{i} B+I&=Q^{-1} \operatorname{diag}\left(\left[\tau_{1}, \cdots, \tau_{d^{2}}\right]\right) Q \\
	&=Q^{-1}\operatorname{diag}\left(N \Sigma_{11}+I_{k}, I_{d^{2}-k} \right)Q,
	\end{aligned}
	\end{equation}
	where $ \tau_1\geq\cdots\geq\tau_{d^2}>0 $, $ \tau_j=O(N) $ for $ 1\leq j\leq k $ and $ \tau_j=1 $ for $ k+1\leq j\leq d^2 $ and the corresponding eigenvectors are $ \{u_j\}_{j=1}^{d^2} $.
	As $ N\rightarrow\infty $, we have
	\begin{equation}
	\begin{aligned}
	&\lim _{N \rightarrow \infty}\left(N S_{i} B+I\right)^{-1}\\
	=&\lim _{N \rightarrow \infty} Q^{-1}\operatorname{diag}\left(N \Sigma_{11}+I_{k}, I_{d^{2}-k} \right)^{-1} Q\\
	=&Q^{-1}\operatorname{diag}\left(0, I_{d^{2}-k} \right) Q,
	\end{aligned}
	\end{equation}
	and thus $ \left(N S_{i} B+I\right)^{-1} $ tends to a constant matrix. Since 
	\begin{equation*}
	I-\left(NS_{i}{B}+ I\right)^{-1}=\left(NS_{i}{B}+ I\right)^{-1}NS_iB,
	\end{equation*}
	it is also a bounded matrix and tends to a constant matrix as $N\rightarrow\infty $. Let the spectral decomposition of $ B $ be 
	\begin{equation}\label{b}
	B=V \Sigma_{2} V^{T}=V\operatorname{diag}\left(\Sigma_{21},0\right) V^{T}.
	\end{equation}
	Thus, the Moore-Penrose inverse of $ B $ is
	\begin{equation}\label{tildeB}
	\tilde{B}=V\operatorname{diag}\left(\Sigma_{21}^{-1},0\right) V^{T},
	\end{equation}
	which is a constant matrix and $ B\tilde{B}B=B $.

	Therefore, the first term of MSE is 
	\begin{equation}
	\begin{aligned}
	&\quad\operatorname{Tr}\left(A_{1}^{-1} A_{2}\right)\\
	&\stackrel{a. s.}{\rightarrow}\operatorname{Tr}\left(\left(N B S_{i}+ I\right)^{-1}\left(N S_{i} B+ I\right)^{-1}  N S_{i} B S_{i}\right) \\
	&=\!\frac{1}{N}\! \operatorname{Tr}\left(\!\left(NS_{i} B+ I\right)^{-1} NS_{i} B\!\cdot\! \tilde{B}\!\cdot\! N{B} S_{i}\left(\!NB S_{i}+ I\right)^{-1}\right)\\
	&= O\left(\frac{1}{N}\right),
	\end{aligned}
	\end{equation}
	because the term  $ \operatorname{Tr}(\cdot) $ is bounded and tends to a constant.
	Therefore, the first term of MSE always scales as $ O(\frac{1}{N}) $. Then we discuss the scaling of the second part of MSE
	\begin{equation}\label{ls2mse}
	\operatorname{Tr}\left(A_{1}^{-1} \theta_{i} \theta_{i}^{T}\right)\stackrel{a. s.}{\rightarrow}\theta_{i}^{T}\left(NBS_i+I\right)^{-1} \left(NS_iB+I\right)^{-1} \theta_{i}.
	\end{equation}
	If $ \theta_{i} $ is a linear combination of $ u_{j}$ for $1 \leq j \leq k   $,  we have
	\begin{equation}
	\operatorname{Tr}\left(A_{1}^{-1}  \theta_{i} \theta_{i}^{T}\right)= O(\frac{1}{N^2}).
	\end{equation}
	Otherwise, if $ \theta_{i} $ is not a linear combination of $ u_{j}, 1 \leq j \leq k   $, $ \operatorname{Tr}\left(A_{1}^{-1}  \theta_{i} \theta_{i}^{T}\right) $ tends to a positive number independent of $ N $.

	Therefore, for the LS MSE $ \mathbb{E}\|\hat E_i- P_i\|^2 $,   it scales as $ O(1/N) $ if and only if the true parameter  $ \theta_{i} $ is a linear
	combination of $  u_{j}$ for $1 \leq j \leq k   $, i.e., $ {\theta}_i\in R(S_{i}{B}) $.  Since $ \{\hat E_i\}_{i=1}^{n} $ may have negative eigenvalues, we use the algorithm in  \cite{wang2019twostage} to further obtain a positive semidefinite  estimate $ \{\hat P_i\}_{i=1}^{n} $.
	The error analysis in \cite{wang2019twostage} has shown that 
	\begin{equation}
	\sum_{i=1}^{n}\left\|\hat{P}_i-P_i\right\|^2=(d n+2 \sqrt{d} n+1) O\left(\sum_{i=1}^{n}\left\|\hat{E}_i-P_i\right\|^2\right).
	\end{equation}
	Therefore, if $  \mathbb{E}\sum_{i=1}^{n}\left\|\hat{E}_i-P_i\right\|^2=O(1/N) $, we have $\mathbb{E}\sum_{i=1}^{n}\left\|\hat{P}_i-P_i\right\|^2=O(1/N)$ and thus the final MSE $\mathbb{E}\left\|\hat{P}_i-P_i\right\|^2$ also scales as $ O(1/N) $. Using \eqref{inv} and \eqref{ls2mse},
	if the true parameter  $ \theta_{i} $ is not the linear
	combination of $  u_{j}$ for $1 \leq j \leq k   $, i.e., $ {\theta}_i\notin R(S_{i}{B}) $, the LS MSE $  \mathbb{E}\left\|\hat{E}_i-P_i\right\|^2 $ tends to a positive value.
	\hfill $\Box$
\end{pf}
\begin{rem}\label{rem22}
In Theorem \ref{theorem1}, when   $\theta_i\notin R(S_{i} B)  $, the behavior of the final MSE $\mathbb{E}\left\|\hat{P}_i-P_i\right\|^2$ is still difficult to characterize. This problem does not exist for a full-rank detector when the resource number $ N $ is large enough, because the LS or WLS estimate already satisfies the positive semidefine constraint and we do not need to correct $ \hat{E}_{i} $.
	
	\end{rem}

Note that when $ S_{i} {B} $ is full-rank, i.e., $ S_{i} $ and $B  $ are both positive definite, the condition  $ \theta_i\in R(S_{i} {B})  $ is always satisfied. Therefore, the MSE always scales as $ O\left(1/N\right) $. Thus, when the types of different probe states are I.C., for any positive definite kernel matrix $ S_i $, the MSE always scales as $ O\left(1/N\right) $. However, when  the probe states are I.I.,	the condition $ \theta_i\in R(S_{i} {B})  $ is difficult to be satisfied  in practice. Thus, without special prior knowledge, for almost all regularization forms, the LS MSE will tend to a constant when $ N $ tends to infinity. In addition, as $ M $ decreases,  for given $ S_i $, this condition may become more difficult to be satisfied because  $ R\left(S_{i} {B}\right) $ may become smaller. Thus,  rank-1 adaptive regularization as in \eqref{first} is not a good choice and full-rank kernel matrix as in \eqref{full} may be better. 
The above analysis can  help understand the boundary of the ability of employing regularization in QDT.

\begin{rem}
	A similar problem was also discussed as Theorem 2.1 in \cite{chentac} for transfer function identification in the I.C. scenario. There a condition to realize unbiased estimation of the true parameters with regularization was given. Here, by allowing the probe states to be I.C. or I.I., we give a stronger result about the scaling of LS MSE as  $ O\left(1/N\right) $ or tends to a constant for QDT.  Our result can also be applied to the case when the variance of noise scales as $ O\left(1/N\right) $, which is typical in the scenario where only statistical noise is considered in quantum measurement.
\end{rem}
\subsection{On the best regularization allowing I.I.}\label{sec52}
We now consider the best regularization which has minimum MSE. It is given by \eqref{best} in the I.C. scenario. Here we aim to characterize the I.I. case. From \eqref{inv} we know $ NS_{i}B+I $ is always invertible.
Define
\begin{equation}
L_{i}\triangleq-\left(N S_{i} B+I\right)^{-1},
\end{equation} and thus
\begin{equation}
I+L_{i}=\left(N S_{i} B+I\right)^{-1} N S_{i} B=-N L_{i} S_{i} B.
\end{equation}
Therefore, we have
\begin{equation}\label{d2b}
\left(I+L_{i}\right) \tilde{B}=
-NL_{i} S_{i} B \tilde{B}.
\end{equation}
Then we propose the following theorem to characterize the best kernel matrix, allowing $ B $ to be singular.
\begin{theorem}\label{theorem2}
	For the $ i $-th POVM element with true parameter $ \theta_{i} $ and  normalized weighted parameterization matrix $ B $ as \eqref{b}, define $\Gamma\triangleq\big\{M \mid M=\theta_{i} \theta_{i}^{T}+V\operatorname{diag}\left(0,Z_{3}\right) V^{T}, Z_{3} \geq 0, \operatorname{dim}(Z_{3})=d^2-\operatorname{rank}(B)\big\}$. If $ \theta_{{i}}\in R(B) $, then $ S_i $ achieves the minimum  of the LS MSE $\mathbb{E}\left\|\hat{E}_i-P_i\right\|^2  $ (i.e., $ S_i $ is the best regularization) if and only if $ S_{i}\in \Gamma$.
	
\end{theorem}
\begin{pf}
	For the LS MSE $\left\|\hat{E}_i-P_i\right\|^2  $ with kernel matrix $ S_i $, using \eqref{A1} and \eqref{A2}, it can be rewritten as
	\begin{equation}\label{mseli}
	\begin{aligned}
	&\operatorname{Tr}\left[\operatorname{MSEM}\mid_{S_{i}}\right]\\
	=&\operatorname{Tr}\left[\left(N S_{i} B+I\right)^{-1}\left(N S_{i} B S_{i}+\theta_{{i}}\theta_{i}^{T}\right)\left(N B S_{i}+I\right)^{-1}\right] \\
	=&\operatorname{Tr}\left[\frac{\left(I+L_{i}\right) \tilde{B}\left(I+L_{i}\right)^{T}}{N}+L_{i} \theta_{{i}}\theta_{i}^{T} L_{i}^{T}\right].
	\end{aligned}
	\end{equation}
	where $ \tilde{B} $ is defined in \eqref{tildeB}.
	Define $g(L_{i})$ to be the last line of (\ref{mseli}).
	Since $ g(L_{i}) $ is convex in $ L_i $, we can find the minimum value by setting the derivative to be zero as
	\begin{equation}\label{grad}
	\frac{dg}{d L_{i}}=\frac{2 \tilde{B}+2 L_{i} \tilde{B}}{N}+2 L_{i} \theta_{{i}}\theta_{i}^{T} =0.
	\end{equation}
	If there exists $S_i\geq0$ so that (\ref{grad}) holds for the corresponding $L_i|_{S_i}$, then such an $S_i$ is the optimal solution to minimize the MSE (\ref{mseli}). We tentatively plug $S_i$ in \eqref{grad}, which (using \eqref{d2b}) becomes  $2 L_{i}\left(-S_{i}B\tilde{B}+\theta_{{i}}\theta_{i}^{T}\right)=0$, equivalent to
	\begin{equation}\label{besteq}
	\tilde{B}BS_{i} =\theta_{{i}}\theta_{i}^{T}.
	\end{equation}
	Since $\theta_{i}\in R(B)  $, we let $\theta_{i}=Bb  $ and then \eqref{besteq} becomes
	\begin{equation}\label{besteq2}
	\begin{aligned}
	&	V\left[\begin{array}{cc}
	I_{21} & 0 \\
	0 & 0
	\end{array}\right] V^{T} S_{i}\\
	=&V\left[\begin{array}{cc}
	\Sigma_{21} & 0 \\
	0 & 0
	\end{array}\right] V^{T} b b^{T} V\left[\begin{array}{cc}
	\Sigma_{21} & 0 \\
	0 & 0
	\end{array}\right] V^{T}.
	\end{aligned}
	\end{equation}
	Denote \begin{equation}
	V^{T} b=\left[\begin{array}{l}
	p \\
	q
	\end{array}\right], V^{T} S_{i} V=\left[\begin{array}{cc}
	Z_{1} & Z_{2} \\
	Z_{2}^{T} & Z_{3}
	\end{array}\right].
	\end{equation}	
	Then \eqref{besteq2} can be simplified as
	\begin{equation}
	\begin{aligned}
	&{\left[\begin{array}{cc}
		I_{21} & 0 \\
		0 & 0
		\end{array}\right]\left[\begin{array}{ll}
		Z_{1} & Z_{2} \\
		Z_{2}^{T} & Z_{3}
		\end{array}\right]=\left[\begin{array}{cc}
		Z_{1} & Z_{2} \\
		0 & 0
		\end{array}\right]} \\
	=&\left[\begin{array}{cc}
	\Sigma_{21} & 0\! \\
	0 & 0
	\end{array}\right]\left[\begin{array}{l}
	p \\
	q
	\end{array}\right]\left[\begin{array}{ll}
	p^{T} & q^{T}
	\end{array}\right]\left[\begin{array}{cc}
	\Sigma_{21} & 0 \\
	0 & 0
	\end{array}\right]\\
	=&\left[\begin{array}{cc}
	\Sigma_{21} p p^{T} \Sigma_{21} & 0 \\
	0 & 0
	\end{array}\right],
	\end{aligned}
	\end{equation}	
	and thus
	\begin{equation}
	Z_{1}=\Sigma_{21} p p^{T} \Sigma_{21}, Z_{2}=0.
	\end{equation}
	Since
	\begin{equation}
	\theta_{i} \theta_{i}^{T}=V\left[\begin{array}{cc}
	\Sigma_{21} p p^{T} \Sigma_{21} & 0 \\
	0 & 0
	\end{array}\right] V^{T},
	\end{equation}
	all solutions  to \eqref{besteq} can be expressed as
	\begin{equation}
	S_{i}\!=\!V\left[\begin{array}{cc}
	\!\Sigma_{21} p p^{T} \Sigma_{21} & 0 \!\\
	\!0 & Z_{3}\!
	\end{array}\right] V^{T}=\theta_{i} \theta_{i}^{T}\!+\!V\left[\begin{array}{cc}
	0 & 0 \\
	0 & Z_{3}
	\end{array}\right] V^{T},
	\end{equation}
	where $ Z_3 $ is positive semidefinite.
	Therefore, the solution set  of \eqref{besteq} is exactly characterized by $ \Gamma $ where
	\begin{equation}
	\begin{aligned}
	\Gamma\triangleq&\big\{M \mid M=\theta_{i} \theta_{i}^{T}+V\operatorname{diag}\left(0,Z_{3}\right) V^{T}, Z_{3} \geq 0,\\
	& \operatorname{dim}(Z_{3})=d^2-\operatorname{rank}(B)\big\}.
	\end{aligned}
	\end{equation}
	\hfill $\Box$
\end{pf}

For all the best regularizations $ S_i $ in $ \Gamma $, we have $ S_iB=\theta_{i} \theta_{i}^{T}B $. This gives the minimum value of the MSE, which can be calculated as
\begin{equation}
\begin{aligned}
&\operatorname{Tr}\left(\operatorname{MSEM}\mid_{S_{i}\in\Gamma}\right)=\operatorname{Tr}\left(\operatorname{MSEM}\mid_{\theta_{i} \theta_{i}^{T}}\right)\\
=&\operatorname{Tr}\bigg[\left(N \theta_{i} \theta_{i}^{T} B+I\right)^{-1}(N \theta_{i} \theta_{i}^{T} B \theta_{i} \theta_{i}^{T}\\
&+\theta_{{i}}\theta_{i}^{T})\left(N B \theta_{i} \theta_{i}^{T}+I\right)^{-1}\bigg] \\
=&\operatorname{Tr}\left[\theta_{{i}}\theta_{i}^{T}\left(N B \theta_{i} \theta_{i}^{T}+I\right)^{-1}\right].
\end{aligned}
\end{equation}
\begin{rem}
Note that the best regularization can minimize  
	$\mathbb{E}\left\|\hat{E}_i-P_i\right\|^2  $ instead of $\mathbb{E}\left\|\hat{P}_i-P_i\right\|^2  $. The question to choose the best regularization to minimize the final MSE $\mathbb{E}\left\|\hat{P}_i-P_i\right\|^2  $ where $ \hat P_i\geq 0 $ is still an open problem. Moreover,
	in practice, we do not know the true values of $ B $ and $ \theta_{{i}} $. One possible solution is to use a rough estimate $ \hat\theta_i $ and $ \hat B_N $ to replace $ \theta_{i} $ and $ B $ in $ \Gamma $. In this case,  there may exist an optimal choice of $ Z_3 $ to achieve the minimum MSE and we leave it as an open problem.
\end{rem}

Here, we compare Theorem \ref{theorem1} and Theorem \ref{theorem2}. If $ \theta_{{i}}\in R(B) $, then for any full-rank kernel matrix $ S_{i} $, $ \theta_{{i}}\in R(S_iB) $ and thus the MSE scales as $O\left(1/N\right) $. For any $ S_i \in \Gamma $, we can obtain the minimum MSE. In addition, $ \theta_{{i}}\in R\left(\theta_{{i}}\theta_{i}^{T}B\right)= R(S_iB) $, and thus the MSE also scales as $O\left(1/N\right) $. If $ \theta_{{i}}\in N(B) $, all the ideal measurement data $ p_{ij} $ are zero, i.e., we cannot obtain any information from the  measurement data. Therefore, $ \theta_{{i}} $ is not identifiable.  If  $ \theta_{{i}}=\theta_{{i,1}}+\theta_{{i,2}} $ where $\theta_{{i,1}}\neq0, \theta_{{i,1}}\in R(B) $ and  $\theta_{{i,2}}\neq 0, \theta_{{i,2}}\in N(B) $, then $ \theta_{{i,1}} $ is identifiable and $ \theta_{{i,2}} $ is not identifiable.
Therefore, we only aim to identify $ \theta_{{i,1}} $ and the discussion is the same as $ \theta_{{i}}\in R(B)  $. 

We then consider two special cases. The first one is that $ B $ is full-rank. Therefore, $ \theta_{i}\in R(B) $ is always satisfied and the unique best kernel matrix  is $ S_i=\theta_{{i}}\theta_{i}^{T} $ which is the same as \cite{CHEN20121525}. The second one is $ S_{i}=\gamma\theta_{{i}}\theta_{i}^{T} $ where $ \gamma $ is a positive constant. Even if $\theta_{{i}}\notin R(B)  $, we still have $ \theta_{{i}}\in R(S_iB) $ ($ \theta_{{i}}\notin N(B) $, otherwise $ R(S_{i}B)=0 $), thus the MSE also scales as $O\left(1/N\right)$.
Note that all the above discussion is based on the assumption that $ N $ tends to infinity. When $ N $ is small, the performance of the regularization forms will be shown through simulation 
in Sec. \ref{ns}.
\section{Numerical simulation}\label{ns}
\begin{figure*}
	\centering
	\includegraphics[width=0.95\textwidth]{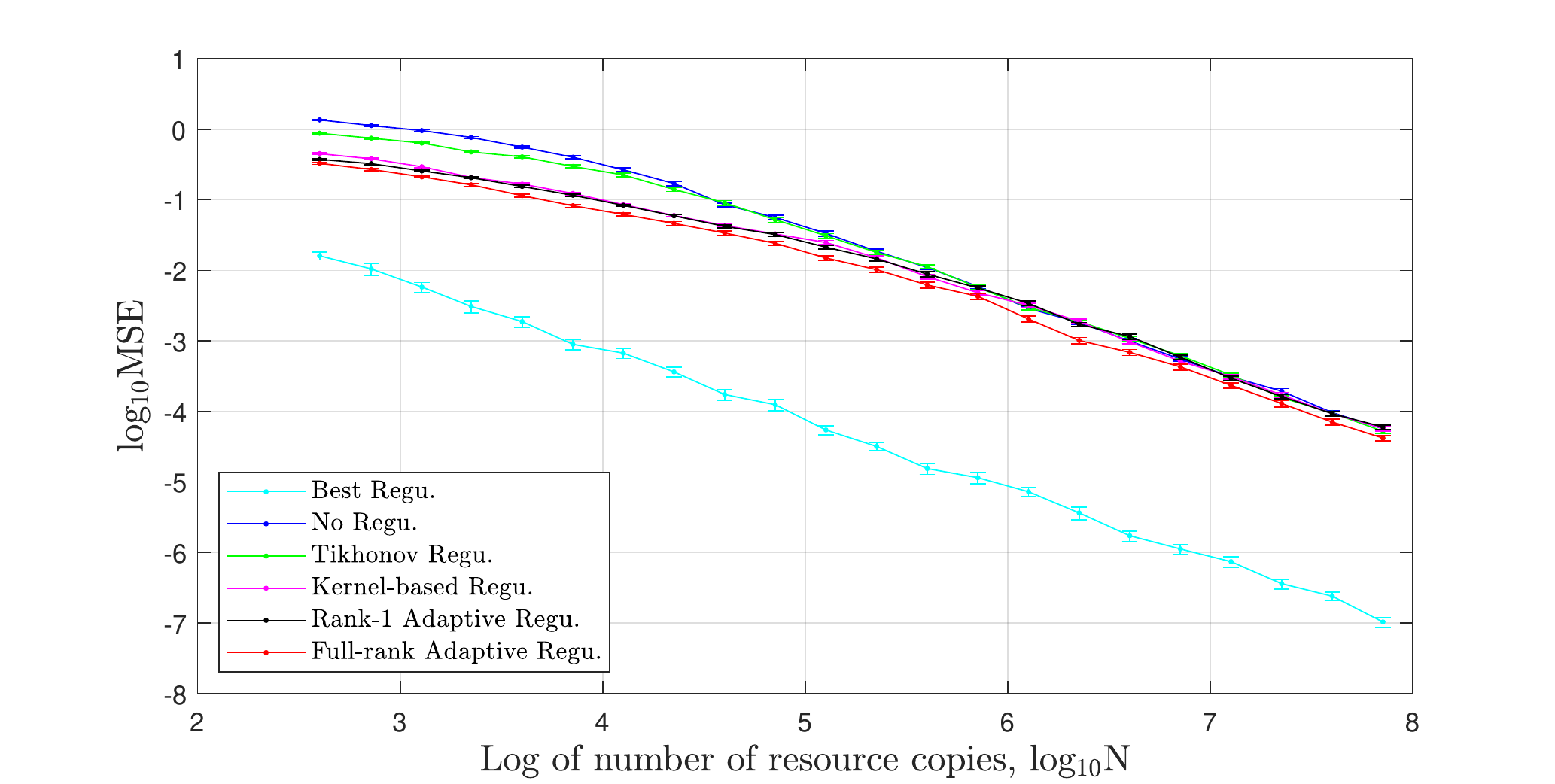}
	\caption{The error scalings of different regularization forms with WLS using $ 20 $   types of $ 4 $ dimensional pure states. When the resource number  $ N>10^6 $, all the MSEs scale as $ O(1/{N}) $ satisfying Theorem \ref{theorem1}. The best regularization is 	$S_i^{\text{best}}=\theta_{i} \theta_{i}^{T}$ which is the lower bound of MSE and  depends on true value of $ \theta_{i} $. Therefore, it cannot be used in practice and we aim to achieve regularization closest to the best regularization. }
	\label{re16}
\end{figure*}
In this section, the evaluation index is the sum of final MSEs $ \mathbb{E}\sum_{i=1}^{n}\left\|\hat{P}_i-P_i\right\|^2 $ and we discuss two commonly used classes of probe states for QDT. The first one involves $ d $ dimensional pure states $ \rho=\left|\psi\right\rangle\left\langle\psi\right|$ where $ \left|\psi\right\rangle$  is the superposition of $ d $ dimensional Fock states as
\begin{equation}\label{pure}
\left|\psi\right\rangle=	\sum_{i=1}^{d} c_{i}\left|i\right\rangle.
\end{equation}
In \cite{xiao2021optimal}, an analysis indicates that pure states may perform better than mixed states for QDT to minimize MSE.

Another class of
commonly used probe states for QDT is the 
coherent states,  because they are more straightforward to be prepared.  A coherent state is denoted as $|\alpha\rangle$ where $\alpha \in \mathbb{C}$ and it can be expanded using
Fock states as
\begin{equation}
|\alpha\rangle=e^{-\frac{|\alpha|^{2}}{2}} \sum_{i=0}^{\infty} \frac{\alpha^{i}}{\sqrt{i !}}|i\rangle.
\end{equation}
Coherent states are in essence infinite dimensional. Denote the corresponding $ d $-dimensional truncation as 
\begin{equation*}
|\alpha_{d}\rangle\triangleq\mathrm{e}^{-\frac{|\alpha|^{2}}{2}} \sum_{i=0}^{d-1} \frac{\alpha^{i}}{\sqrt{i} !}|i\rangle.
\end{equation*}
To estimate a $d$ dimensional detector, in the simulation we assume that the outcomes generated by the residual signal $ \operatorname{Tr}\left[\left(|\alpha\rangle-\left|\alpha_{d}\right\rangle\right)\left(\langle\alpha|-\left\langle\alpha_{d}\right|\right)\right] $  are all included in the outcomes of the last POVM element. Since we truncate the coherent state in $ d $-dimension, $ \operatorname{Tr}\left(\left|\alpha_{d}\right\rangle\langle\alpha_{d}|\right)<1$ but for pure states in \eqref{pure} $ \operatorname{Tr}\left(\rho\right)=1 $. Here we discuss resource distribution optimization without regularization and
different regularization forms under the uniformly distributed resources.

When applying kernel-based regularization, an important problem is to determine the hyper-parameters (such as $c$, $ \mu $ in \eqref{DI}, \eqref{tc} and \eqref{dc}) in the kernel matrix $ S_i $. In this paper, we apply the same kernel matrix for all the POVM elements and use cross-validation in \cite{CHEN20121525} to determine these hyper-parameters:
	\begin{enumerate}
		\item[(1)] Split the probe states randomly into two parts: an estimation data part with probe state parameterization matrix $ X_1 $ and a validation data part with probe state parameterization matrix $ X_2 $.
		\item[(2)] Collect all the hyper-parameters in a vector $ \omega $. Then 
		estimate the detector as $ \bar{\theta}_i $ using the measurement data from $ X_1 $ for different candidate values of hyper-parameters $\omega \in \bar{\Omega} $ where $ \bar{\Omega} $ is a finite set in our paper.
		\item[(3)] Using the validation data from $ X_2 $, we  find
		\begin{equation}
		\omega_{0}=\arg\min_{\omega\in \bar{\Omega}} \sum_{i=1}^{n}\|\hat{y}_i-X_{2}\bar{\theta}_i(\omega)\|^2.
		\end{equation}
		The model can then be re-estimated for this $\omega_0$ using all the probe states. Other methods to determine the hyper-parameters can also be found in \cite{CHEN20121525,Chen2013}.
	\end{enumerate}

\subsection{Superposed Fock states}\label{purenum}
We consider a $ 4$ dimensional three-valued phase-sensitive detector, which is close to phase-insensitive detector as 
\begin{equation}\label{detector4}
\begin{aligned}
P_{1}^{(4)}&\!\!=\!\!\left[\begin{array}{cccc}\!\!0.1 & \!\!0 & \!0.002-0.005 \mathrm{i} & 0.003+0.007 \mathrm{i} \!\!\\ \!\!0 &\!\! 0.2\!\! & 0 & 0 \!\!\\ \!\!0.002+0.005 \mathrm{i} & \!\!0\!\! & 0.3 & 0 \!\!\\ \!\!0.003-0.007 \mathrm{i} & \!\!0\!\! & 0 & 0.4\!\!\end{array}\right],\\
P_{2}^{(4)}&\!\!=\!\!\left[\begin{array}{cccc}
0.2 & 0.001+0.002 \mathrm{i} & 0 & 0 \\
0.001-0.002 \mathrm{i} & 0.2 & 0 & 0 \\
0 & 0 & 0.3 & 0 \\
0 & 0 & 0 & 0.4
\end{array}\right],\\
P_{3}^{(4)}&=\!\!I-P_{1}^{(4)}-P_{2}^{(4)}.
\end{aligned}
\end{equation}

Using the  algorithm in \cite{MISZCZAK2012118,qetlab}, we generate  $ 20 $ different types of $ 4 $ dimensional pure states.
To determine the hyper-parameters in the DI kernel, we use $16$ pure states as estimation data and $ 4 $ pure states as validation data.
We use different regularization forms including no regularization (\eqref{Ti} with $ c=0$), Tikhonov regularization (\eqref{Ti} with $ c=10$), kernel-based regularization (\eqref{DI} with $c=0.1, \mu=0.9 $), rank-1 adaptive regularization,  full-rank  adaptive regularization (see Sec. \ref{sec5})  and the best regularization \eqref{best}. The best regularization is the lower bound of MSE and  depends on true value of $ \theta $. Therefore, it cannot be used in practice and we aim to achieve regularization closest to the best regularization.  For rank-1 adaptive regularization, we use kernel-based regularization (\eqref{DI} with $c=0.1, \mu=0.9 $) in step 1 and   \eqref{first} in step 2. For full-rank  adaptive regularization, we use kernel-based regularization (\eqref{DI} with $c=0.1, \mu=0.9 $) in step 1 and   \eqref{full} in step 2. For each resource number, we run the algorithm $ 100 $ times and obtain the average MSE and standard deviation.

The results are shown in Fig. \ref{re16}. The best regularization scales as $ O(1/{N}) $ satisfying Theorem \ref{theorem1}. When the resource number  $ N<10^6 $, the MSEs of kernel-based regularization and  adaptive regularization are a little smaller than Tikhonov regularization and no regularization. In addition, full-rank adaptive regularization has a little smaller MSE than rank-1 adaptive regularization. When the resource number  $ N>10^6 $, all the MSEs scale as $ O(1/{N}) $ satisfying Theorem \ref{theorem1}.

\begin{figure}
	\centering
	\includegraphics[width=0.5\textwidth]{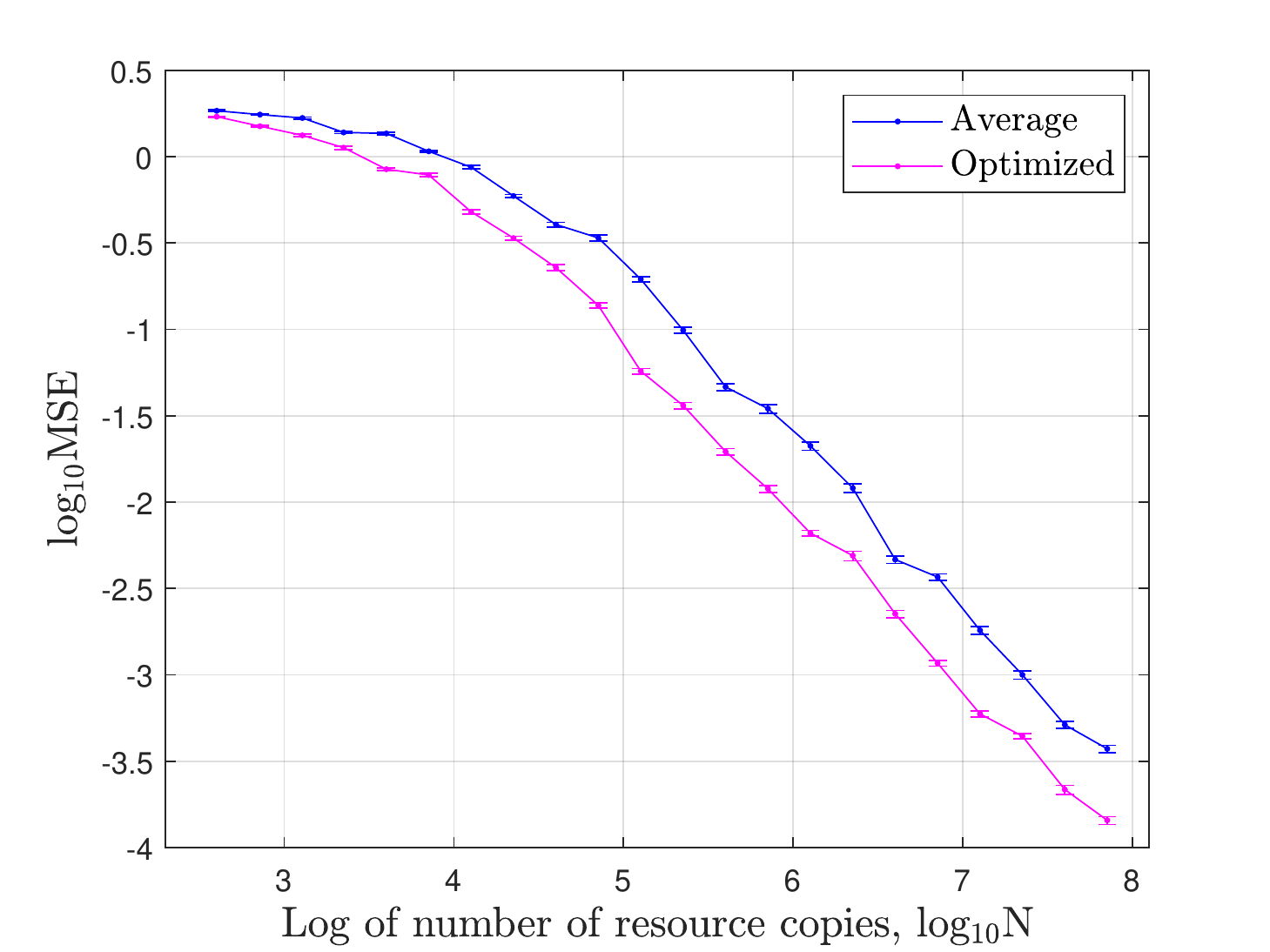}
	\caption{The MSE comparison between average and optimized 
		resource  distribution  using $ 20 $   types of $ 4 $ dimensional pure states.}
	\label{rsop}
\end{figure}

Since these $ 4 $ dimensional pure states are I.C., without regularization, we also consider resource distribution optimization. We compare the MSE of the case with averagely distributed resources $N/M$ (``Average" in Fig. \ref{rsop}) and the MSE of the case with optimized resource distribution (``Optimized" in Fig. \ref{rsop}) by solving \eqref{rs2}.
For each resource number $ N $, we run the algorithm $ 100 $ times and obtain the average MSE and standard deviation. The results are shown in Fig. \ref{rsop}. We can obtain a lower MSE with resource  distribution optimization and both MSEs scale as $ O\left(1/{N}\right) $ when $ N>10^5 $.

\begin{figure}
	\centering
	\includegraphics[width=0.5\textwidth]{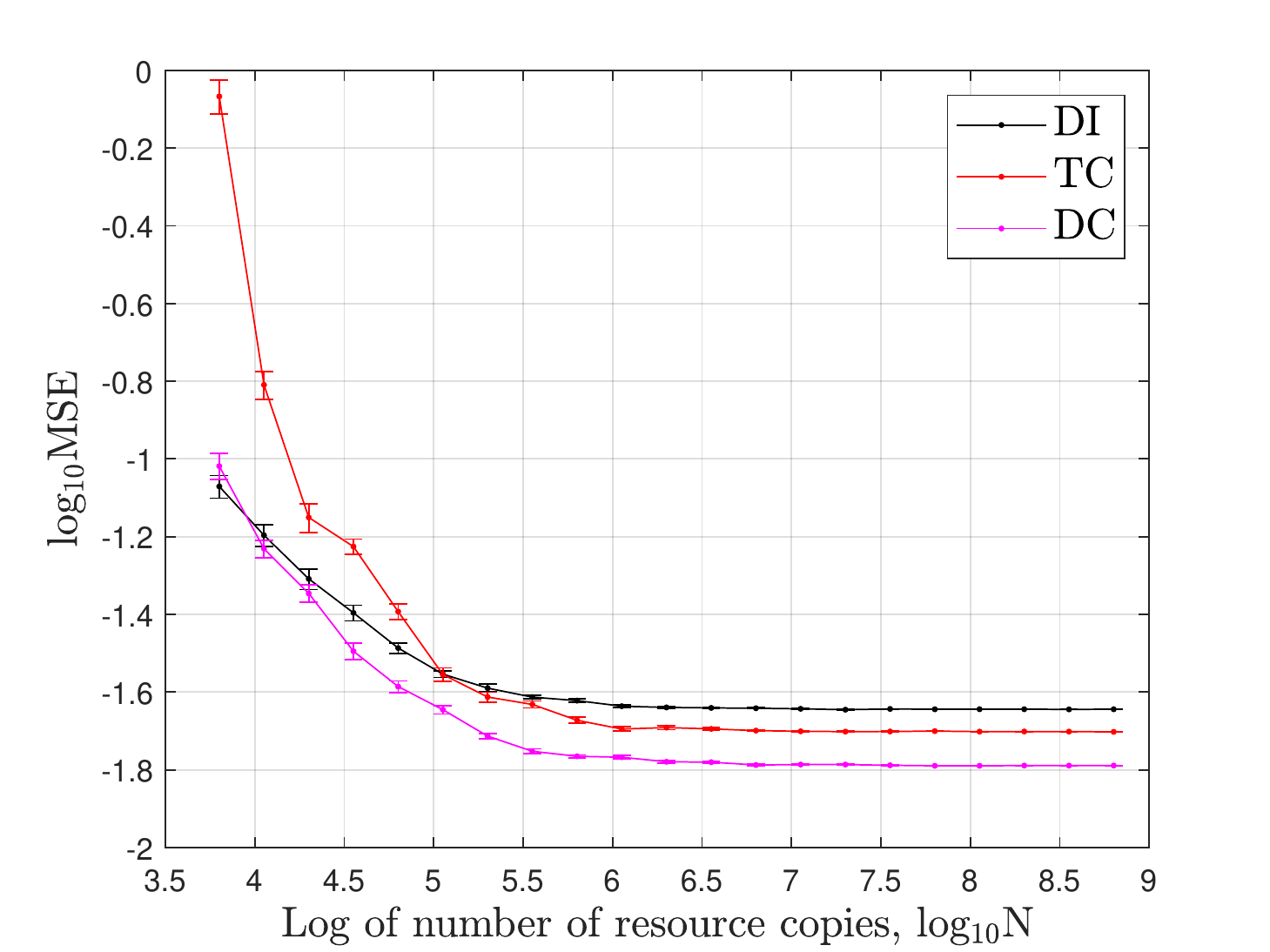}
	\caption{The MSE comparison of DI, TC  and DC kernels with WLS using $ 10 $  types of $ 4 $ dimensional pure states. All the  MSEs tend to  constants as predicted by Theorem \ref{theorem1} and Remark \ref{rem22} because $ \theta_{i}\in R(S_{i}B) $ does not hold.} 
	\label{kernel}
\end{figure}

Then we generate only $ 10 $ random types of $ 4 $ dimensional pure states. To determine the hyper-parameters in the different kernels, we use $8$ pure states as the estimation data and $ 2 $ pure states as the validation data.
Here we assume that we have the prior knowledge that the true detector is close to a phase-insensitive detector. We choose Pauli basis \begin{equation}
\Omega=\left\{I \otimes I, I \otimes \sigma_z, \sigma_z \otimes I, \sigma_z \otimes \sigma_z, \cdots, \sigma_x \otimes \sigma_y\right\} / 2.
\end{equation}
Thus, the absolute values of the first four elements in $ \theta_{{i}} $ are significantly larger than zero and all the other values in $ \theta_{{i}} $ are close to zero, which is similar to the impulse responses of stable transfer functions in the system identification. Therefore, we  use  DI kernel (\eqref{DI} with $c=0.1, \mu=0.9 $), TC kernel (\eqref{tc} with $c=0.9, \mu=0.8 $) and DC kernel (\eqref{dc} with $c=0.1, \mu_1=0.2, \mu_2=0.9 $), and compare their performance.  The results are shown in Fig. \ref{kernel}. Compared  with DI kernel, the MSE of DC kernel is improved by $28.3\%$ when $ N>10^{6} $, which indicates that DC kernel is suitable to be applied for calibrating phase-insensitive detectors.

\begin{figure*}
	\centering
	\includegraphics[width=0.9\textwidth]{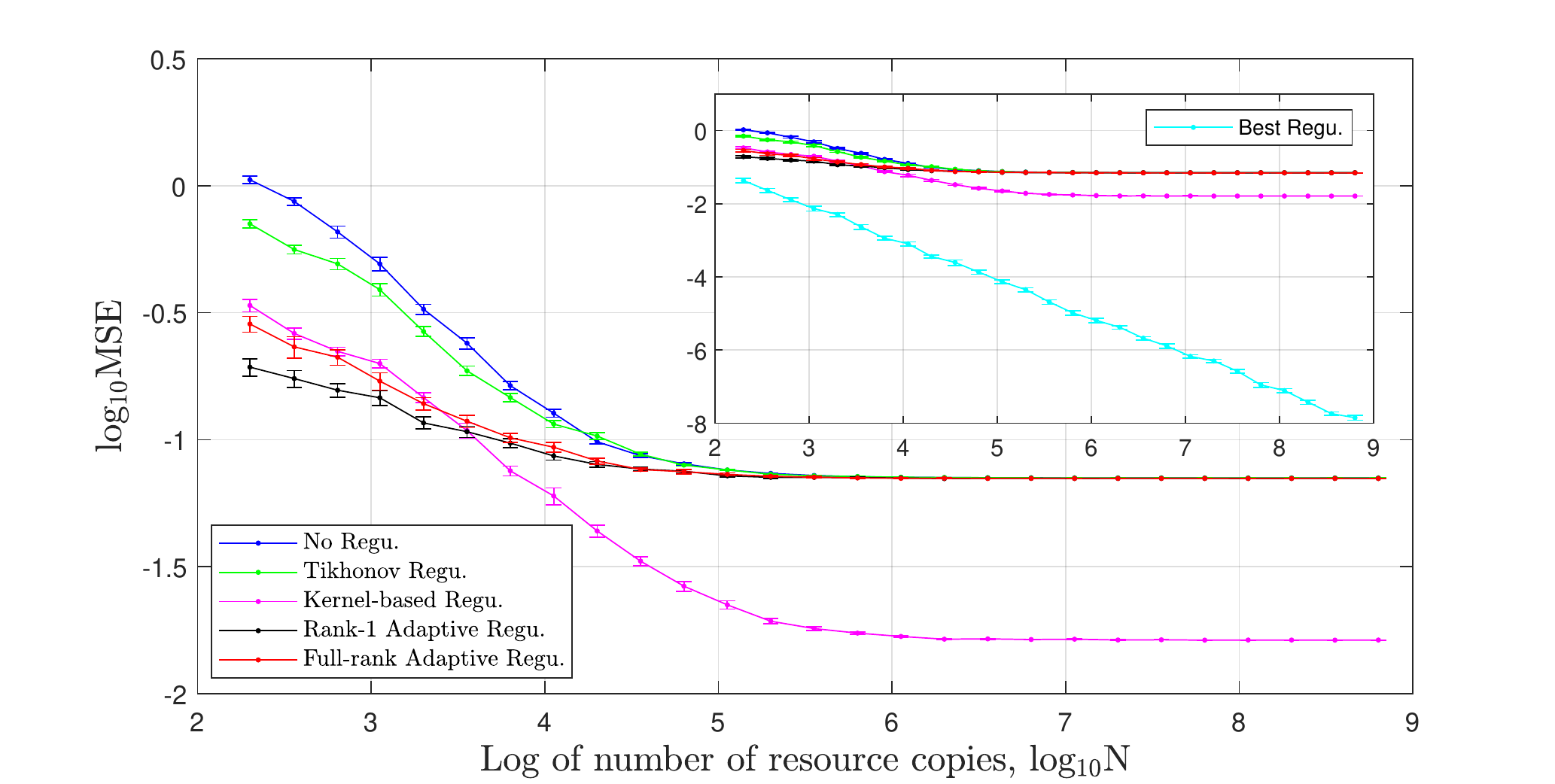}
	\caption{The error scalings of different regularization forms with WLS using $ 10 $  types of $ 4 $ dimensional pure states. Except the best regularization, all the  MSEs tend to  constants as predicted by Theorem \ref{theorem1} and Remark \ref{rem22} because $ \theta_{i}\in R(S_{i}B) $ does not hold. Using true parameters $ \theta_{i} $, the best regularization is 	$S_i^{\text{best}}=\theta_{i} \theta_{i}^{T}$ and thus $ \theta_{i}\in R(S_{i}B) $ always holds. According to Theorem \ref{theorem1}, the best regularization scales as $ O\left(1/{N}\right) $ for arbitrary detectors.}
	\label{re10}
\end{figure*}

Hence we change DI kernel to DC kernel (\eqref{dc} with $c=0.1, \mu_1=0.2, \mu_2=0.9 $) in this case and
 the results are shown in Fig. \ref{re10}.  In this I.I. scenario, there does not exist a unique solution for WLS \eqref{awls} without regularization. Therefore, we use the  Moore-Penrose inverse of $\tilde X_{i}^{T}\tilde X_{i}  $ to obtain an estimate instead of \eqref{awls}, which is called ``no regularization" in Fig. \ref{re10}. 
The best regularization also scales as $ O(1/{N}) $ satisfying Theorem \ref{theorem1}.
Kernel-based regularization has the minimum MSE compared with other regularization forms because DC kernel utilizes the prior knowledge on the sparsity of coefficients. In addition, the MSEs of adaptive regularizations are always a little smaller than Tikhonov regularization and no regularization when $ N<10^{5} $. 

Here we explain the reason why adaptive regularization with rank-1 kernel matrix fails to exhibit a clear advantage over typical non-adaptive protocol (as shown in Fig. \ref{re10}) in the I.I. scenario.
In  the first step, for the chosen kernel matrix $ S_{i}^{(1)} $, the condition
$ {\theta}_i\in R(S_{i}^{(1)}{B}) $ is usually not satisfied in the I.I. scenario. Thus, the estimate $ \hat{\theta}_{i}^{0} $ is biased and MSE tends to a positive constant $ c $ as 
\begin{equation}\label{adaptiveerror}
\lim _{N \rightarrow \infty} \mathbb{E}\left\|\theta_{i}- \hat{\theta}_{i}^{0}\right\|=c>0.
\end{equation}
Then in  the second step, if we choose regularization as \eqref{first}, $ {\theta}_i\notin R(S^{\text{rank-1}}_{i}{B}) $  because the only one vector in $ R(S^{\text{rank-1}}_{i} B) $ is $ {\hat\theta}_{i}^{0} $ and $	\lim _{N \rightarrow \infty} \mathbb{E}\left\|\theta_{i}- \hat{\theta}_{i}^{0}\right\|=c>0 $. Moreover, even if we use multi-step regularization with rank-1 kernel matrix as above, the estimation result is still biased and MSE always tends to a constant, because the number of adaptive steps is always finite.  
As $ N $ increases, except the best regularization, all the  MSEs tend to  constants as predicted by Theorem \ref{theorem1} because $ \theta_{i}\in R(S_{i}B) $ does not hold.

\subsection{Coherent states}
Since coherent states are truncated, we consider a larger dimensional  three-valued phase-sensitive detector as
\begin{equation}
\begin{aligned}
&P_{1}^{(8)}=U_{1} \operatorname{diag}\left(P_{1}^{(4)}, P_{1}^{(4)}\right) U_{1}^{\dagger}, \\
&P_{2}^{(8)}=U_{2} \operatorname{diag}\left(P_{2}^{(4)}, P_{2}^{(4)}\right) U_{2}^{\dagger},\\
&P_{3}^{(8)}=I-P_{1}^{(8)}-P_{2}^{(8)},
\end{aligned}
\end{equation}
where $ d=8 $ and $ U_1 $, $ U_2 $ are random unitary matrices \cite{Zyczkowski_1994,qetlab}. We also ensure $ P_{3}^{(8)} $ is positive semidefinite.

Since coherent states are more similar to each other, we generate $ 640 $ random different types of coherent states using the probe state preparation in \cite{wang2019twostage} where the real part and imaginary part of $ \alpha $ are randomly generated in the interval $ [-1,1] $.
We use different regularization forms including no regularization (\eqref{Ti} with $ c=0$), Tikhonov regularization (\eqref{Ti} with $ c=10$), kernel-based regularization (\eqref{DI} with $c=0.2, \mu=0.9 $), rank-1 adaptive regularization, full-rank  adaptive regularization (see Sec. \ref{sec5})  and the best regularization \eqref{best}. For rank-1 adaptive regularization, we use kernel-based regularization (\eqref{DI} with $c=0.2, \mu=0.9 $) in step 1 and   \eqref{first} in step 2. For full-rank  adaptive regularization, we use kernel-based regularization (\eqref{DI} with $c=0.2, \mu=0.9 $) in step 1 and   \eqref{full} in step 2. For each resource number, we run the algorithm $ 100 $ times and obtain the average MSE and standard deviation.

\begin{figure*}
	\centering
	\includegraphics[width=0.95\textwidth]{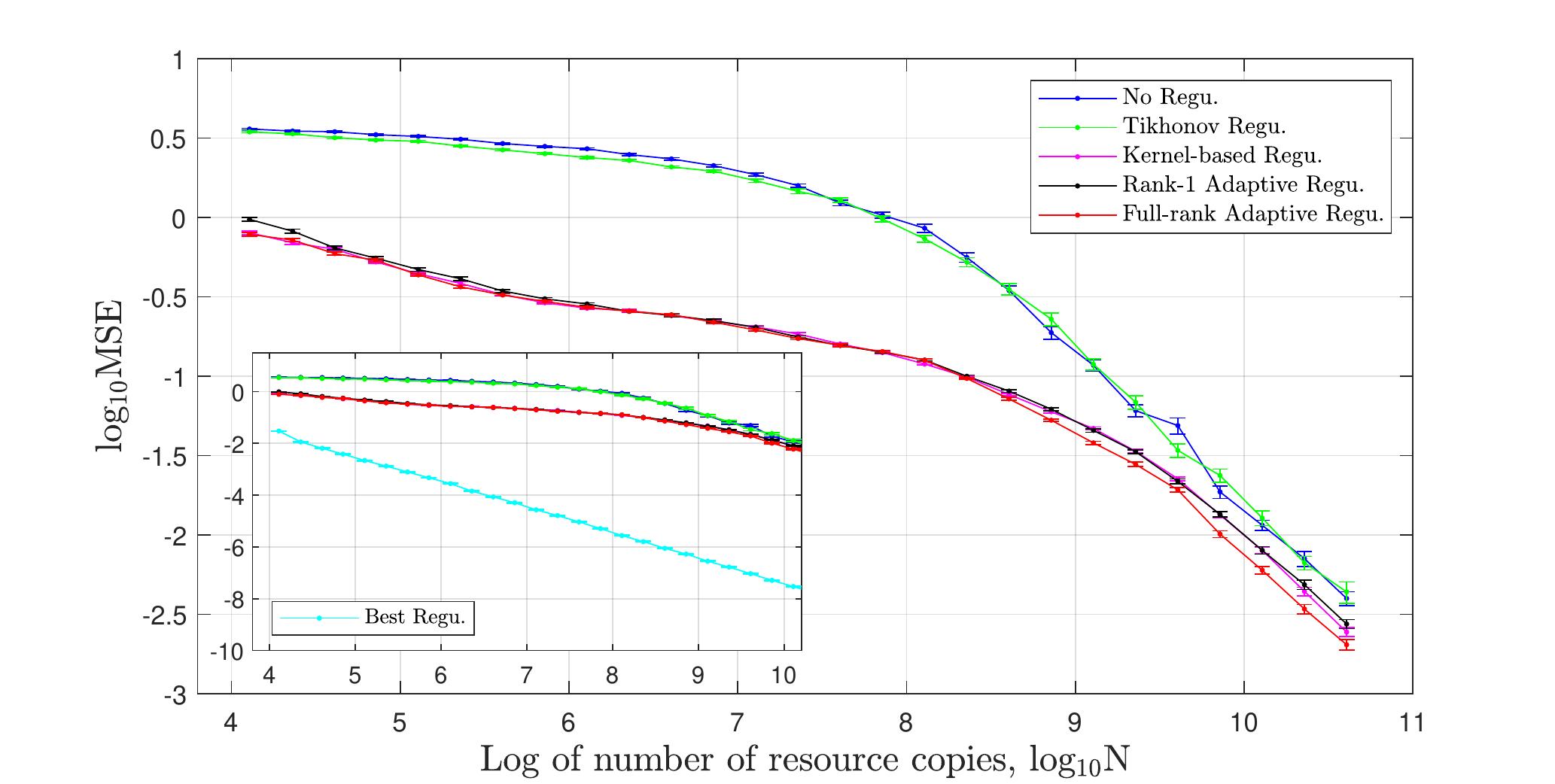}
	\caption{The error scalings of different regularization forms with WLS using $ 640 $  types of coherent states.  When $ N>10^{8} $, all the MSE scales as $ O(1/{N}) $ satisfying Theorem \ref{theorem1}. The best regularization is 	$S_i^{\text{best}}=\theta_{i} \theta_{i}^{T}$ which is the lower bound of MSE and  depends on true value of $ \theta_{i} $. Therefore, it cannot be used in practice and we aim to achieve regularization closest to the best regularization.}
	\label{co640}
\end{figure*}

\begin{figure}
	\centering
	\includegraphics[width=0.5\textwidth]{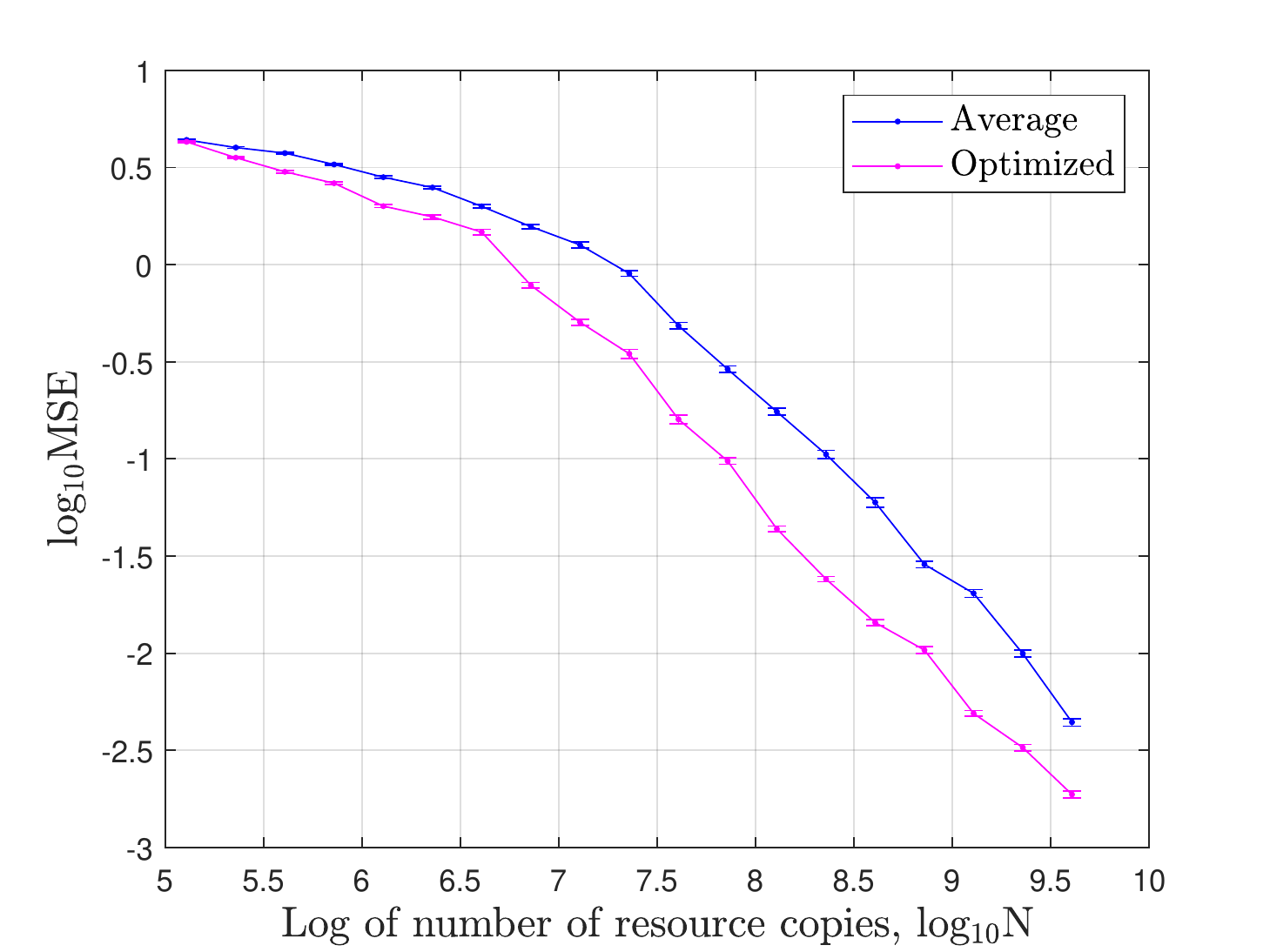}
	\caption{The MSE comparison between average and optimized 
		resource  distribution using $ 640 $ types of coherent states.}
	\label{co640op}
\end{figure}

\begin{figure*}
	\centering
	\includegraphics[width=0.95\textwidth]{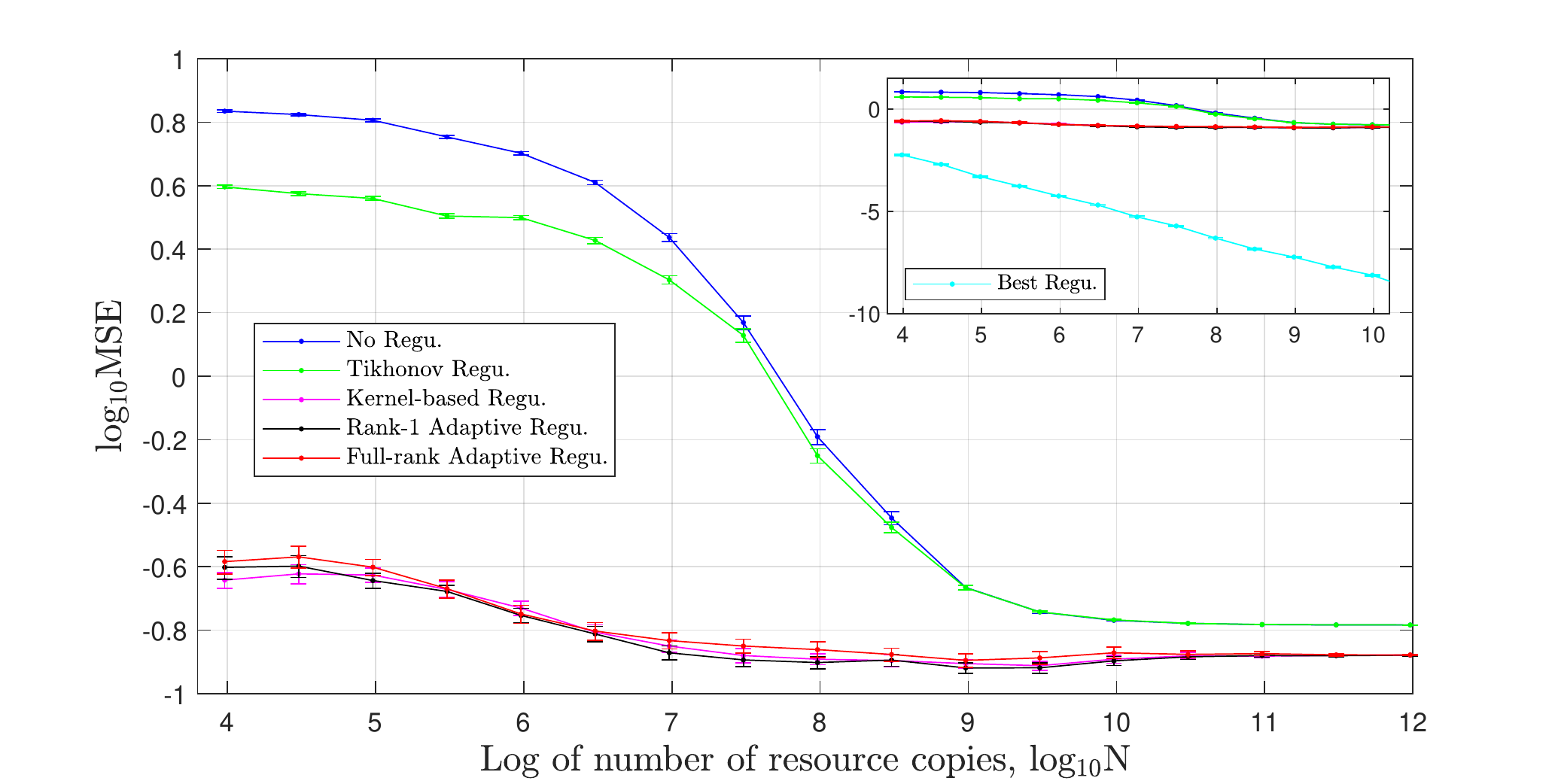}
	\caption{The error scalings of different regularization forms with WLS using $ 48 $  types of coherent states. Except the best regularization, all the  MSEs tend to  constants as predicted by Theorem \ref{theorem1} and Remark \ref{rem22} because $ \theta_{i}\in R(S_{i}B) $ does not hold. Using true parameters $ \theta_{i} $, the best regularization is 	$S_i^{\text{best}}=\theta_{i} \theta_{i}^{T}$ and thus $ \theta_{i}\in R(S_{i}B) $ always holds. According to Theorem \ref{theorem1}, the best regularization scales as $ O\left(\frac{1}{N}\right) $ for arbitrary detectors.}
	\label{co48}
\end{figure*}

 The results are shown in Fig. \ref{co640}. When   $ N<10^8 $, the MSEs of kernel-based regularization and  adaptive regularization are a little smaller than Tikhonov regularization and no regularization. In addition, full-rank adaptive regularization has a little smaller MSE than rank-1 adaptive regularization. When $ N>10^{8} $, all the MSEs scale as $ O(1/{N}) $ satisfying Theorem \ref{theorem1}. Since these coherent states are I.C., we also consider resource distribution optimization without regularization. The simulation results are shown in Fig. \ref{co640op}. We can also obtain a lower MSE with resource distribution optimization and both MSEs scale as $ O\left(1/{N}\right) $ for $ N>10^{7} $. Then using the  same algorithm in \cite{wang2019twostage}, we generate only $ 48 $ random types of coherent states where real parts and imaginary parts of $ \alpha $ are randomly generated in the interval $ [-1,1] $.
 We use the same regularization \eqref{DI} and the results are shown in Fig. \ref{co48}.
 Kernel-based regularization and adaptive regularization always have smaller MSEs compared with Tikhonov regularization and no regularization. When $ N>10^{10} $, except the best regularization, all the  MSEs tend to  constants as predicted by Theorem \ref{theorem1}  because $ \theta_{i}\in R(S_{i}B) $ does not hold.

\section{Experimental examples}\label{secexp}
We consider the same quantum optical experimental system for QDT in \cite{Yokoyama:19} and \cite{wang2019twostage}.
Ref. \cite{wang2019twostage} used Tikhonov regularization based on standard LS to complete the QDT. Here, we consider the same experimental data and employ kernel-based regularization based on WLS instead to further improve the QDT accuracy.
\subsection{Experimental setup}\label{expset}
\begin{figure}
	\centering	\includegraphics[width=3.3in]{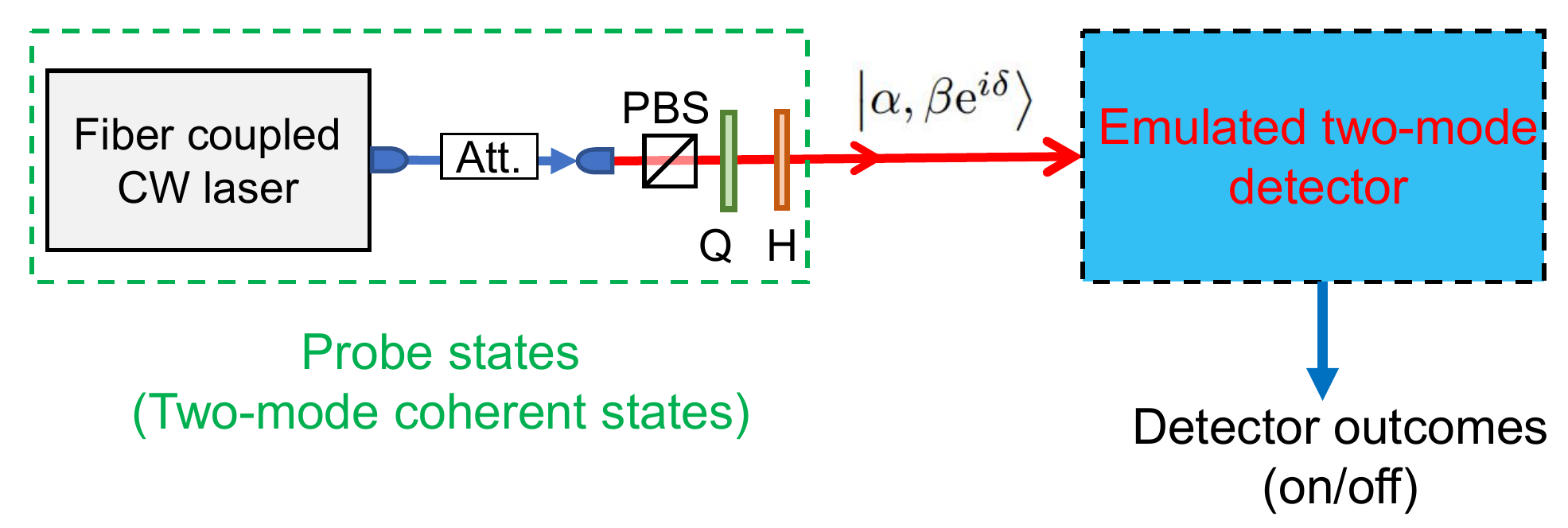}
	\caption{The detailed structure and description for 	experimental setup can be found in \cite{wang2019twostage,Yokoyama:19}.
		Att., Attenuator;
		PBS, Polarization beam splitter;
		H, Half wave plate;
		Q, Quarter wave plate.
		In experiment, we generate two-mode coherent states as \eqref{twomode} and input them to emulated two-mode detector. Then we obtain detector outcomes and identify this two-mode detector. 
	}
	\label{fig}
\end{figure}

The entire experimental setup is given in Fig. \ref{fig}, which determines the structure of the detector to be estimated. More details about this setup can be found in \cite{wang2019twostage,Yokoyama:19}. It leads to block-diagonal binary detectors $ P_0+P_1=I $ as
\begin{equation}\label{eqa43}
P_i=L^{(i)}_1\oplus L^{(i)}_2\oplus \cdots\oplus L^{(i)}_m,
\end{equation}
where $m$ is the number of different blocks and $L^{(i)}_j\geq 0$ is $d_j\times d_j$ dimensional, with $\sum_{j=1}^m d_j=d$. Hence, we need to identify each block $L^{(i)}_j$. 
Two-mode coherent states are prepared for detector tomography by using an adequately attenuated continuous-wave (CW) fiber coupled laser as depicted in the green dashed box in Fig. \ref{fig} \cite{wang2019twostage,Yokoyama:19}.
We express the general two-mode coherent state without global phase as $|\alpha,\beta \text{e}^{\text{i}\delta}\rangle$ ($\delta\in\mathbb{R}$, $\alpha,\beta\geq0$), which can be expanded in the Fock state basis as
\begin{equation}\label{twomode}
|\alpha,\beta \text{e}^{\text{i}\delta}\rangle=\exp[-\frac{1}{2}(\alpha^2+\beta^2)] \sum_{j,k}^{\infty}\frac{\alpha^j\beta^k\text{e}^{\text{i}k\delta}} {\sqrt{j!k!}}|j,k\rangle,
\end{equation}
and the parameters of the $ 19 $ probe states used are shown in \cite{wang2019twostage,Yokoyama:19}. The amplitudes of these coherent states satisfy $(\alpha, \beta) \!\!\in\!\!\{\!(0.316,0.316),\!(0.447,0),\!(0,0.447),\!(0.194,0.112),$\\$(0.112,0.194), (0,0)\}$.

Although the probe states are I.C., the condition number of the probe states' parameterization matrix $ X $ is large and the problem is ill-conditioned. Thus, we add regularization to identify each block $ L_j $. 
After regularized WLS, we obtain an estimate $\{\hat E_i\}$ which might not be positive semidefinite. Then we use the Stage 2 algorithm as in \cite{wang2019twostage} in each block and obtain $ \hat Q^{(i)}_j $.
The final estimation is thus $\hat P_i=\hat Q^{(i)}_1\oplus \hat Q^{(i)}_2\oplus  \cdots\oplus \hat Q^{(i)}_m$, which is physical and also satisfies the block-diagonal requirement.

\subsection{Result comparison}
Ref. \cite{wang2019twostage} considered experiments for two different sets of detectors, denoted as Group I and Group II, respectively, and  the basis of the POVM elements is the two-mode
Fock state basis as $ \{|0,0\rangle,|1,0\rangle,|0,1\rangle,|2,0\rangle,|1,1\rangle,|0,2\rangle\}$. For the true value of Group I, $P_1=L_1^{(1)}\oplus L_2^{(1)}\oplus L_3^{(1)}$, and we have $L_1^{(1)}=2.91\times 10^{-4}$,
\begin{equation}
L_2^{(1)}=\left[
\begin{array}{cc}
0.202&0.00109\text{i}\\
-0.00109\text{i}&0.202\\
\end{array}\right],\nonumber
\end{equation}
and
\begin{equation}
L_3^{(1)}=\left[
\begin{array}{ccc}
0.363&0.00123\text{i}&1.20\times10^{-6}\\
-0.00123\text{i}&0.363&0.00123\text{i}\\
1.20\times10^{-6}&-0.00123\text{i}&0.363\\
\end{array}\right].\nonumber
\end{equation}
For the true value of Group II, we have $L_1^{(1)}=1.27\times 10^{-4}$,
\begin{equation}
L_2^{(1)}=\left[
\begin{array}{cc}
0.0763&-0.0440+0.0879\text{i}\\
-0.0440-0.0879\text{i}&0.127\\
\end{array}\right],\nonumber
\end{equation}
and $L_3^{(1)}=$
\begin{equation}
\!\!\left[\!\!
\setlength{\arraycolsep}{1.6pt}
\begin{array}{ccc}
0.147&-0.0574+0.115\text{i}&0.00580+0.00773\text{i}\\
-0.0574-0.115\text{i}&0.184&-0.0543+0.109\text{i}\\
0.00580-0.00773\text{i}&-0.0543-0.109\text{i}&0.238\\
\end{array}\!\right]\!\!\!.\nonumber
\end{equation}

\begin{figure}
	\centering
	\includegraphics[width=0.48\textwidth]{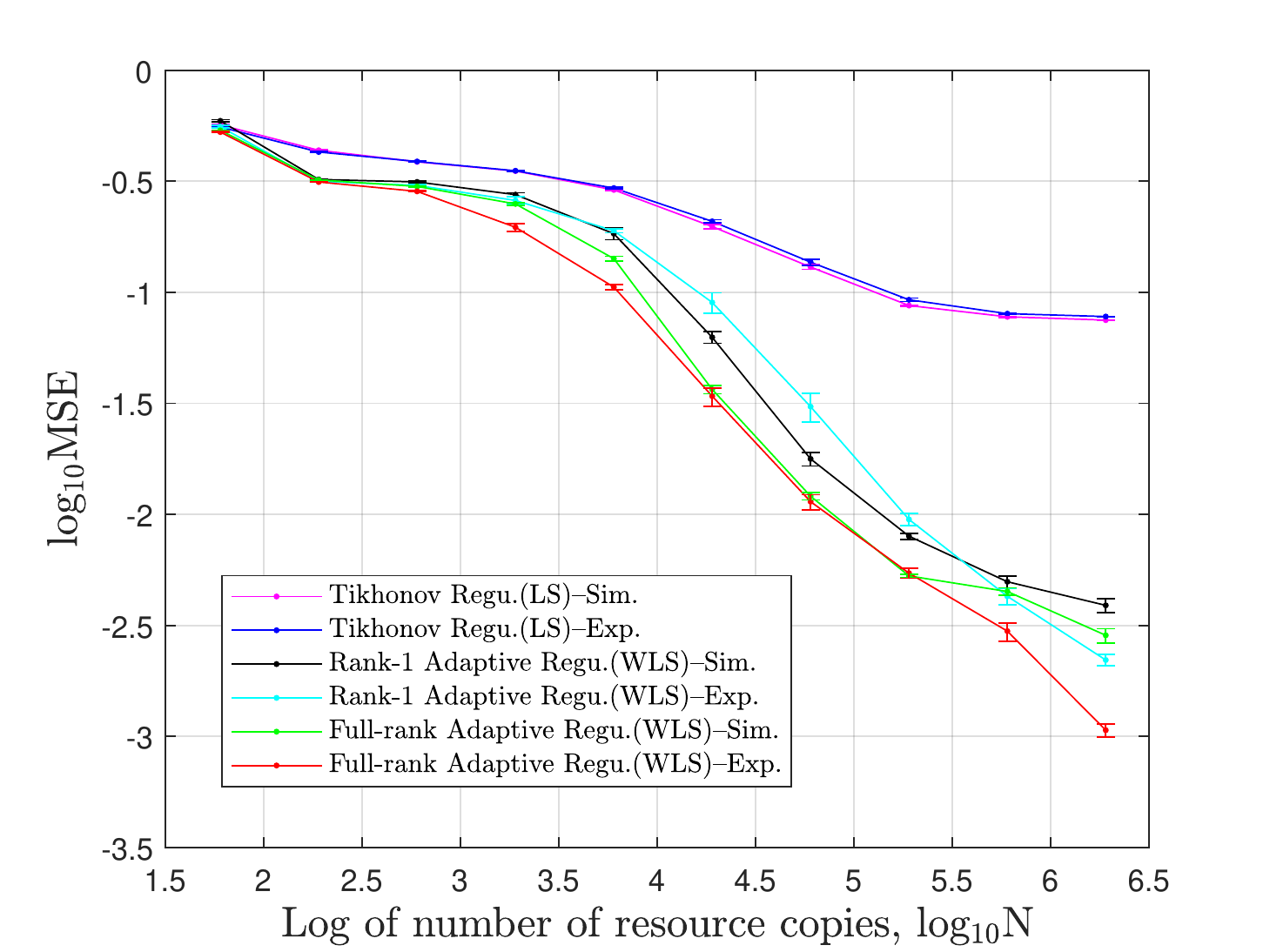}
	\caption{Experimental and simulation QDT results of Tikhonov regularization (LS), rank-1 adaptive regularization (WLS) and full-rank adaptive regularization (WLS)  for Group I.}
	\label{exp1}
\end{figure}
\begin{figure}
	\centering
	\includegraphics[width=0.48\textwidth]{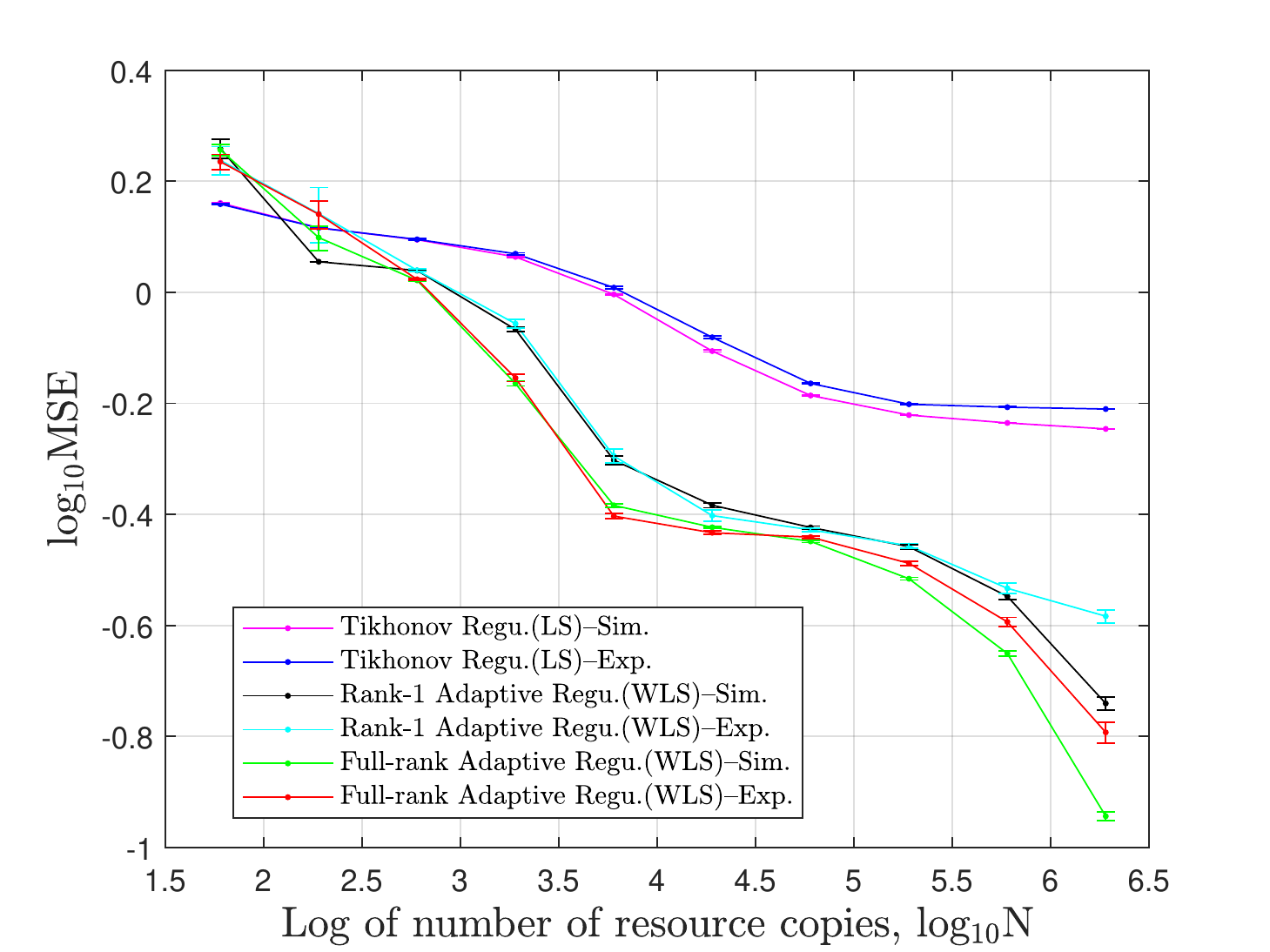}
	\caption{Experimental and simulation QDT results of Tikhonov regularization (LS), rank-1 adaptive regularization (WLS) and full-rank adaptive regularization (WLS) for Group II.}
	\label{exp2}
\end{figure}
Ref. \cite{wang2019twostage} recorded $10^{6}$ measurement outcomes for each input state, and repeated the process $6$ times.  We use these measurement data to identify the detectors and  also plot the identification results using simulated measurement data as a comparison in Fig. \ref{exp1} and Fig. \ref{exp2}.

For the QDT problem,  Ref. \cite{wang2019twostage} employed Tikhonov regularization with standard LS estimation, where they chose $ D_i^{\text{Tikhonov}}=\frac{10^3}{N} I $ and the estimation is given in \eqref{wangtik}, while here we use rank-1 adaptive regularization and full-rank adaptive regularization  with WLS. Since the result of kernel-based regularization is similar to adaptive regularization, we only show the results of adaptive regularization.

To determine the hyper-parameters in the DI kernel, we use $15$ probe states as estimation data and $ 4 $ probe states as validation data.
In Group I,  we choose $c=0.001, \mu=0.8 $ in \eqref{DI} in step 1 and \eqref{first} in step 2 for-rank 1 adaptive regularization and for full-rank  adaptive regularization, we choose $c=0.001, \mu=0.8 $ in \eqref{DI} in step 1 and \eqref{full} in step 2. The results are shown in Fig. \ref{exp1}.
Adaptive regularization (WLS) performs  better than  Tikhonov regularization (LS) in \cite{wang2019twostage}, especially for large resource number $ N $. In addition, the MSE of full-rank adaptive regularization is a little smaller than rank-1  adaptive regularization.
In Group II, for rank-1 adaptive regularization, we choose $c=0.0008, \mu=0.9 $ in \eqref{DI} in step 1 and \eqref{first} in step 2 and for full-rank  adaptive regularization, we choose $c=0.0008, \mu=0.9 $ in \eqref{DI} in step 1 and \eqref{full} in step 2.  The results are shown in Fig. \ref{exp2}. Adaptive regularization (WLS) performs  better than  Tikhonov regularization (LS) when $ N>10^{2.5} $ and the MSE of full-rank adaptive regularization is always a little smaller than rank-1  adaptive regularization. Moreover,
the MSE of Group II is a little larger than that of Group I because  the amplitudes of nondiagonal elements in Group II are significantly larger than zero.
\section{Conclusion}\label{con}
In this paper, using regularization, we improve QDT accuracy with  given probe states. In the I.C. and I.I. scenarios, we have employed WLS estimation, discussed different regularization forms, proved the scaling of MSE under the static assumption and characterized the best regularization. In the I.C. scenario, without regularization, we have studied resource distribution optimization and converted it to an SDP problem.
The numerical examples have demonstrated the effectiveness of different regularization forms and resource  distribution optimization. In a quantum optical experiment, our adaptive regularization with WLS has achieved lower mean squared errors compared with Tikhonov regularization with LS.  It remains an open problem how to choose the kernel optimally in adaptive regularization.

\begin{ack}
	S. Xiao would like to thank Dr. Xueke Zheng for the helpful discussion.
\end{ack}

\bibliographystyle{ieeetr}         
\bibliography{redetector}

\end{document}